\documentclass[final]{siamltex}

\usepackage{epsf}
\usepackage{amssymb}
\usepackage{amsbsy}
\usepackage{verbatim}

\usepackage{graphicx}

\usepackage{datetime}

\usepackage[colorlinks=true, linkcolor=blue, citecolor=blue, filecolor=blue, runcolor=blue, urlcolor = blue]{hyperref}

\newcommand{\OpFldDissp}{\mu\Delta}

\newcommand{\OpFSCouple}{\Gamma}
\newcommand{\OpSFCouple}{\Lambda}

\newcommand{\OpFldDisspDiscr}{L}

\newcommand{\OpFSCoupleDiscr}{\Gamma}
\newcommand{\OpSFCoupleDiscr}{\Lambda}

\newcommand{\OpAdjoint}{{\dagger}}
\newcommand{\RegAdjoint}{{T}}

\newcommand{\Eng}{\Phi}

\newcommand{\mb}[1]{\mathbf{#1}}

\newcommand{\subtxt}[1]{ {\mbox{\tiny #1}} }

\begin{document}

\title{Incorporating Shear into Stochastic Eulerian Lagrangian Methods
for Rheological Studies of Complex Fluids and Soft Materials
}

\author{Paul J. Atzberger
\thanks{University of California, 
Department of Mathematics , Santa Barbara, CA 93106; 
e-mail: atzberg@math.ucsb.edu; phone: 805-893-3239;
Work supported by NSF DMS-0635535 and NSF CAREER Grant DMS - 0956210.
\url{http://atzberger.org/}
}
}

\maketitle

\begin{abstract}
We develop computational methods that incorporate shear into fluctuating hydrodynamics methods.  We are motivated by the rheological responses of complex fluids and soft materials.  Our approach is based on continuum stochastic hydrodynamic equations that are subject to shear boundary conditions on the unit periodic cell in a manner similar to the Lees-Edwards conditions of molecular dynamics.  Our methods take into account consistently 
the microstructure elastic mechanics, fluid-structure hydrodynamic 
coupling, and thermal fluctuations.  For practical simulations, we develop numerical methods for efficient stochastic field generation that handle the sheared generalized periodic boundary conditions.  We show that our numerical methods are consistent with fluctuation dissipation balance and near-equilibrium statistical mechanics.  As a demonstration in practice, we present several prototype rheological response studies.  These include (i) shear thinning of a polymeric fluid, (ii) complex moduli for the oscillatory responses of a polymerized lipid vesicle, and (iii) aging under shear of a gel-like material.   
\end{abstract}

\begin{keywords}
Statistical Mechanics, Complex Fluids, Soft Materials, 
Stochastic Eulerian Lagrangian Methods, SELM, Stochastic Immersed 
Boundary Methods, SIB, Fluctuating Hydrodynamics, Fluid-Structure 
Coupling, Polymeric Fluid, FENE, Vesicles, Gels. 
\end{keywords}

\noindent 
\small
Related Software: 
\url{https://github.com/atzberg/mango-selm}.\\ 
Additional Information: 
\url{http://atzberger.org/}.
\\
 
\pagestyle{myheadings}
\thispagestyle{plain}
\markboth{P.J. ATZBERGER}
{SIMULATION OF COMPLEX FLUIDS AND SOFT MATERIALS USING SELM}

\section{Introduction} 
In the study of complex fluids and soft materials an important aim is to understand 
how macroscopic properties emerge from microstructure order and kinetics.  
Examples include liquid crystals,
colloidal suspensions, gels, lipids, and 
emulsions~\cite{Lubensky1997, Hamley2003, bird1987, 
Gompper2006, Doi1986, Coussot2007, Raub2007}.  
The microstructures of these materials often have interactions on energy scales 
comparable to thermal energy resulting in phases dependent on the balance between 
enthalpic and entropic effects~\cite{Larson1999,Doi1986,Bird1987Vol2}.  
For biological materials, such as the cell membrane or cytoskeleton, 
microstructures may in addition exert their own active forces~\cite{Alberts2002}.  
The individual and collective dynamics of the microstructures often span a wide range of length scales and time 
scales~\cite{Gompper2006,Larson1999}.  To obtain insights, simplified models are often developed which are
tailored to specific mechanistic questions about material responses.  To perform simulations of dynamic responses 
requires tractable numerical methods that can account for microstructure mechanics, hydrodynamics, 
and the roles played by thermal fluctuations~\cite{Danuser2006,Li2005,Mizuno2007}.

We present an approach based on the Stochastic Eulerian Lagrangian Method (SELM)~\cite{AtzbergerSELM2011}.
In SELM the microstructure mechanics and solvent dynamics is formulated at the level of continuum mechanics.
The fluid-structure interactions are treated on an approximate level avoiding the need for computations of the surface traction stresses or 
coupling tensors.  The SELM approach is closely related to the Stochastic Immersed Boundary  
Method~\cite{Atzberger2007a, Peskin2002}, Force Coupling Method~\cite{Maxey2001}, Stokesian-Brownian Dynamics~\cite{Brady1988, Banchio2003}, Arbitrary Eulerian-Lagrangian Method~\cite{Braescu2007}, Fluctuating Hydrodynamics Methods of~\cite{DeFabritiis2007a, Vazquez-Quesada2009, BalboaUsabiaga2012, Donev2009a}, and others~\cite{Cortez2001, VazquezQuesada2012, Fujita2008,Ottinger1997a,Ottinger1997b, Dunweg2007}.  The SELM formulation can be used to study results in different limiting physical regimes either incorporating or neglecting various types of coupling and inertial effects~\cite{AtzbergerSELM2011, AtzbergerSELMRed2013}.  In this work, we extend SELM for rheological studies in the following ways:
(i) we formulate generalized periodic boundary conditions for the fluctuating hydrodynamic equations that account for shear deformations, 
(ii) we develop numerical discretizations for the fluid equations based on a moving reference frame to handle the shear deformation and boundary conditions, and  
(iii) we develop efficient computational methods for stochastic field generation methods to account for thermal fluctuations.  We also perform analysis of the time-dependent numerical discretizations to study the statistical mechanics of the numerical methods.  

The SELM framework is presented in Section~\ref{sec_SELM}.  To extend SELM to incorporate shear, 
we introduce generalized periodic boundary conditions in 
Section~\ref{sec_SELM_shear_bnd}.  We develop stochastic numerical 
methods for the fluctuating hydrodynamic equations in Section~\ref{sec_comp_method}.  To demonstrate 
how the methods work in practice, we present results for a few example
applications in Section~\ref{sec_applications}.  We present results for the 
shear thinning of a polymeric fluid
in Section~\ref{sec_appl_shear_thinning}.  
We investigate the complex moduli 
for the oscillatory responses of a polymerized lipid 
vesicle in Section~\ref{sec_appl_osc_responses}.  We study the 
aging of the shear viscosity of a gel-like material in 
Section~\ref{sec_appl_aging_shear_viscosity}.  Overall, we expect 
the presented approaches to be useful in adopting fluctuating 
hydrodynamics descriptions to investigate diverse models and 
phenomena arising in studies of complex fluids and soft materials.

\section{Stochastic Eulerian Lagrangian Method}
\label{sec_SELM}
We describe the microstructure and solvent dynamics using the Stochastic Eulerian Lagrangian Method (SELM)~\cite{AtzbergerSELM2011}.
This has the fluid-structure equations
\begin{eqnarray}
\label{equ_SELM_III_1}
\rho
\frac{d\mathbf{u}}{dt}  & = & \OpFldDissp \mathbf{u} -\nabla{p} + \OpSFCouple[-\nabla_{\mathbf{X}}\Phi(\mathbf{X})] 
                         + \left(\nabla_{\mathbf{X}}\cdot\OpSFCouple\right) k_B{T}
                         + \mathbf{g}_\subtxt{thm} \\
\nabla\cdot\mb{u}       & = & 0 \\
\label{equ_SELM_III_2}
\frac{d\mathbf{X}}{dt}  & = & \OpFSCouple\mathbf{u} \\
\label{equ_SELM_III_3}
\langle \mathbf{g}_\subtxt{thm}(s)\mathbf{g}_\subtxt{thm}^{\RegAdjoint}(t) \rangle & = & -\left(2k_B{T}\right)\OpFldDissp\hspace{0.06cm}\delta(t - s).
\end{eqnarray}
The $\mb{u}$ denotes the fluid velocity and $\mb{X}$ the microstructure configurations.
The potential energy of the microstructures is given by $\Phi[\mb{X}]$.  It is assumed throughout 
that to a good approximation the solvent fluid can be treated as incompressible with Newtonian stresses~\cite{Acheson1990, Bird1987Vol1}.  
The $\mb{u}$ denotes the fluid velocity, $\rho$ denotes the uniform fluid density, $\mu$ the dynamic fluid viscosity, 
and $p$ the fluid pressure.  To account for thermal fluctuations,
we introduce a stochastic driving field $\mb{g}_\subtxt{thm}$ which is assumed to be a 
Gaussian process with mean zero and $\delta$-correlation in time.  The notation 
$\mb{a}(s)\mb{b}^T(t)$ should be interpreted as the tensor product of $\mb{a}$ and $\mb{b}$.  The transpose notation 
is used to be consistent with the discrete setting and the outer-product between column vectors.  

\begin{figure}[t]
\centering
\includegraphics[width=5in]{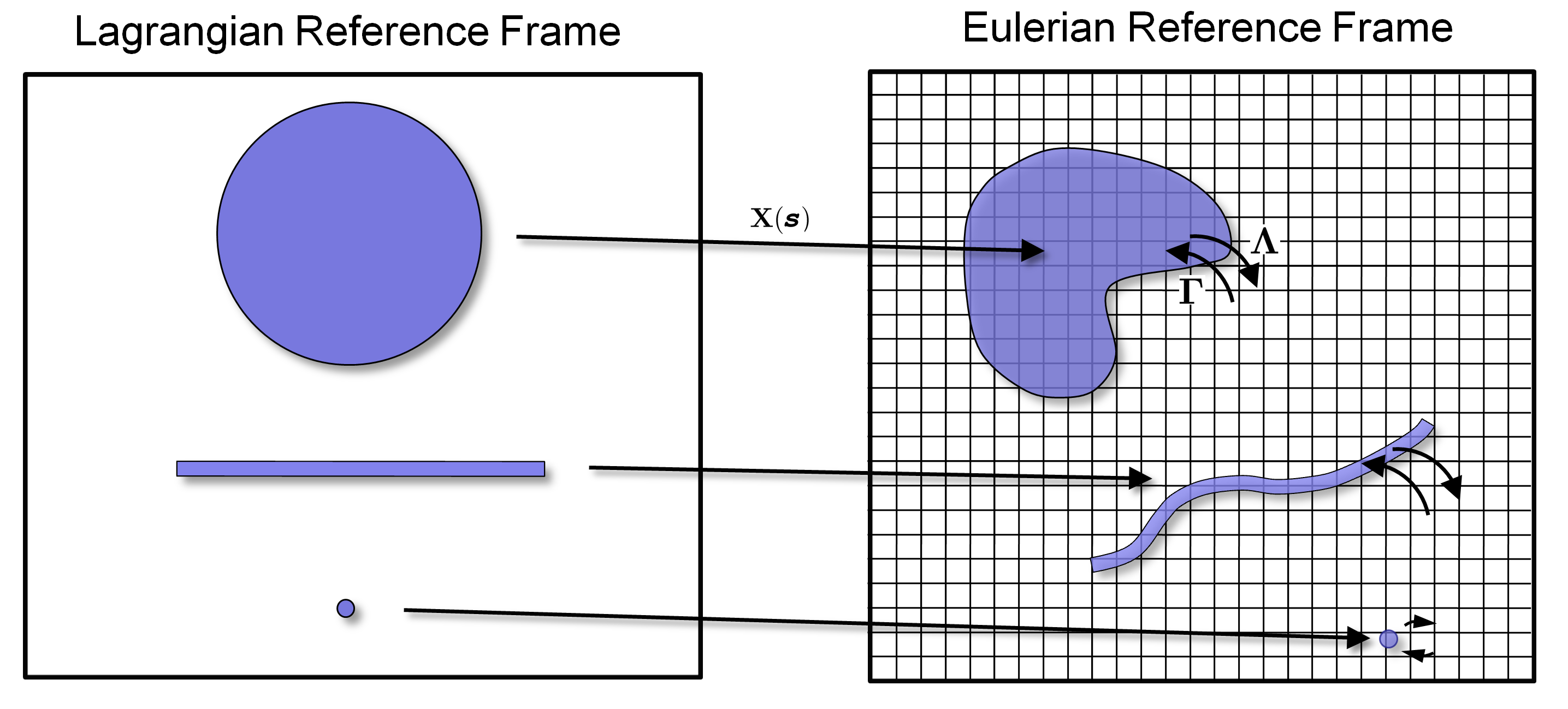}
\caption[Mixed Eulerian and Lagrangian Description]
{The description of the fluid-structure system utilizes both
Eulerian and Lagrangian reference frames.  The structure mechanics
are often most naturally described using a Lagrangian reference 
frame.  The fluid mechanics are often most naturally described
using an Eulerian reference frame.  The mapping $\mathbf{X}(\mathbf{s})$ 
relates the Lagrangian reference frame to the Eulerian reference frame.  
The operator $\OpFSCouple$ prescribes how structures are to be 
coupled to the fluid.  The operator $\OpSFCouple$ prescribes how the 
fluid is to be coupled to the structures.  A variety of fluid-structure
interactions can be represented in this way.  This includes rigid and 
deformable bodies, membrane structures, polymeric structures, or point 
particles.  
}
\label{figure_EL_schematic}
\end{figure}

The fluid and microstructure degrees of freedom are coupled through the linear operators $\OpFSCouple$, $\OpSFCouple$, see Figure~\ref{figure_EL_schematic}.  The operators themselves are assumed to have dependence only on the configuration degrees 
of freedom and time $\OpFSCouple = \OpFSCouple[\mb{X},t]$, $\OpSFCouple = \OpSFCouple[\mb{X},t]$.
To ensure the coupling is non-dissipative and conserves energy, the 
following adjoint condition is imposed~\cite{AtzbergerSELM2011,Peskin2002}
\begin{eqnarray}
\label{equ_adjoint_cond}
\int_{\mathcal{S}} (\OpFSCouple\mb{u})(\mb{s}) \cdot \mathbf{v}(\mb{s}) d\mb{s} 
= 
\int_{\Omega} \mb{u}(\mb{x}) \cdot (\OpSFCouple\mathbf{v})(\mb{x}) d\mb{x}.
\end{eqnarray}
This is required to hold for any $\mathbf{u}$ and $\mathbf{v}$.  The adjoint condition ensures 
the fluid-structure coupling is not a source of energy dissipation and provides a model having 
properties similar to imposing a no-slip boundary condition on the microstructures.  This
condition is also important to the fluctuation-dissipation balance in the system and simplifies
the formulation by ensuring there is no need for additional stochastic driving fields to 
compensate for losses in the 
fluid-structure coupling.  The $\mathcal{S}$ and $\Omega$ 
denote the spaces used to parameterize respectively 
the microstructure configurations and the fluid.  We denote adjoints in the sense of equation~\ref{equ_adjoint_cond} by
$\OpSFCouple = \OpFSCouple^{\OpAdjoint}$ and $\OpFSCouple = \OpSFCouple^{\OpAdjoint}$.
A specific form for the coupling operators will be given in Section~\ref{sec_operators}.  

We remark that SELM can also accomodate active forces exerted on the microstructures which would appear in the fluid equations in the same place as $-\nabla \Phi$ term.  An important constraint given the periodic unit cell used in simulations is that the total momentum must be conserved.   This is assumed when solving the fluid equations to determine the degenerate constant mode of the fluid (usually one assumes the total momentum is zero).   If this is not the case, then artefacts in which there is a global “back-flow” of the fluid can arise in simulations to compensate for the unbalanced forces acting on the system.  Provided the active forces are introduced in a balanced manner the fluid-structure formulation and numerical methods we present can be used.

It should be mentioned that when interpreting the fluctuating hydrodynamic equations the thermal fluctuations cause significant 
irregularity in the fluid velocity field $\mb{u}$.
In fact, the $\mb{u}$ is not defined in a point-wise sense as a classical function but only in the sense of 
a generalized function (distribution)~\cite{Lieb2001}.  As a consequence, some care must be taken in the treatment of the material
derivative $d\mb{u}/dt = \partial \mb{u}/\partial t + \mb{u}\cdot \nabla{\mb{u}}$, see~\cite{AtzbergerSELM2011, Donev2009a}.  
The convective term involves a product of distribtions and is not mathematically well-defined.  The convective term 
arises from deriving a local statement of the conservation of momentum as a differential equation in the Eulerian 
frame of reference.  The issue has to do with the tacit assumption in continuum mechanics that deformations of the 
material body are smooth which does not hold in this stochastic setting.  A number of ways to handle this issue can be considered.  
One is to introduce a regularisation length-scale to smooth the velocity field motivated on physical grounds
by the fact that the hydrodynamic description is not expected to hold below sufficiently small length-scales since 
the continuum hypothesis breaks down as we approach the mean-free path of the fluid molecules.  Provided this
regularisation preserves the skew-symmetry of the convective term, it would not contribute to dissipatation of energy and 
would not change signiciantly the stochastic driving fields and numerical methods we present.  From our dimension analysis 
of the fluctuating hydrodynamics equations, it would seem that the time derivative term plays the more dominate role given
the rapid oscillations introduced by the thermal fluctuations, see discussion in~\cite{AtzbergerSELM2011,AtzbergerSELMRed2013}.  However, 
such comparisons can be subtle since the time-derivative by itself is also not a well-defined term and requires an interpretation 
be given to the SPDEs, such as Ito stochastic calculus~\cite{Oksendal2000}.  To avoid these technical issues, we consider here only models 
that use the time-dependent Stokes equations with the linearized material derivative $d\mb{u}/dt = \partial \mb{u}/\partial t$.

\section{Extension of SELM for Investigations of Shear Responses}
\label{sec_SELM_shear_bnd}

\begin{figure}[t]
\centering
\includegraphics[width=5in]{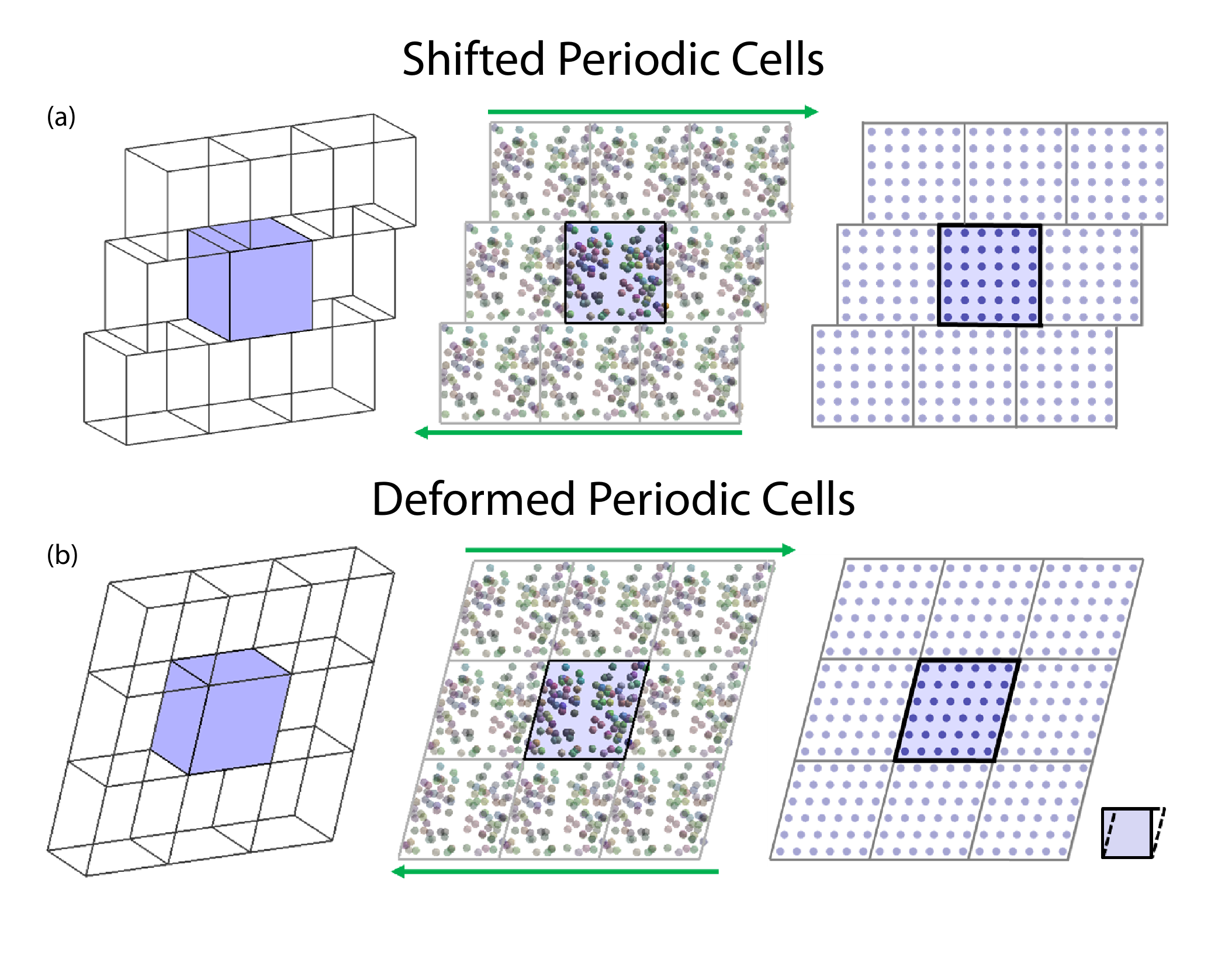}
\caption[Lees-Edwards Boundary Conditions and Discretization Mesh]
{Lees-Edwards Boundary Conditions and Discretization Mesh.
The boundary conditions can be thought of in two equivalent ways.  The first is 
to consider shear induced by shifting the periodic images of
the unit cell by a shift relative to the unit cell, see panel (a).  
If the unit cell is discretized directly using a standard uniform mesh this presents 
challenges since the mesh sites become misaligned at boundaries between the unit cell 
and the periodic images.  The second is to use the periodic symmetry 
which allows for the shifted images to be equivalently expressed in terms of a periodic
tiling of a deformed unit cell, see panel (b).  By discretizing the momentum equations in a 
moving coordinate frame, a discretization mesh is obtained which aligns with the periodic images.  
In this case, the unit cell and mesh change over time from a cube to a sheared parallelepiped.
The periodicity further introduces a symmetry by which a unit cell distorted
by the shift $\frac{1}{2}L$ is equivalent to the distortion by shift $-\frac{1}{2}L$.  Using
this symmetry we always keep in the numerical calculations the unit cell deformation 
within the range $[-\frac{1}{2}L,\frac{1}{2}L]$.
}
\label{figure_meshLeesEdwardsI}
\label{figure_meshLeesEdwardsII}
\end{figure}

To introduce shear we generalize the usual periodic 
boundary conditions.  We are motivated by
the approach introduced by Lees-Edwards for 
molecular dynamics methods~\cite{LeesEdwards1972,Evans1979,EvansMorriss1984}.
In this work, the material is modeled by periodically repeating
the unit cell of a molecular model.  To simulate the material
undergoing a shear deformation at a given rate, the 
periodic images are treated as shifting in time relative
to the unit cell.
This has the effect of modifying both the location of periodic 
images of molecules and their assigned velocities.  This 
approach has some advantages over methods
which enforce a strict affine-like deformation everywhere within 
the material body~\cite{EvansMorriss1990,Hoover1980,Hoover2008}.  
For the Lees-Edwards approach, the shear is imposed only at 
the boundaries allowing within the unit cell for the molecular 
interactions to determine the shear response, see 
Figure~\ref{figure_meshLeesEdwardsI}.  

Motivated by this approach, we develop a corresponding methodology 
for the SELM approach.  By considering the effect of shifting 
periodic images of the unit cell, we introduce the generalized 
periodic boundary conditions for the fluid velocity
\begin{eqnarray}
\label{equ_StokesJumpBndCond}
\mathbf{u}(x,y,L,t) = \mathbf{u}(x - vt,y,0,t) + v\mathbf{e}_{x}.
\end{eqnarray}
To simplify the presentation, we only consider the case where a shear 
is imposed in the z-direction giving shear induced velocities in the 
x-direction.  The other cases follow similarly.  In our notation, the 
$L$ is the side length of the periodic cell in the z-direction,
$v = L\dot{\gamma}$ is the velocity of the top face of the unit cell 
relative to the bottom face,  $\dot{\gamma}$ denotes the rate of 
shear deformation, and $\mathbf{e}_j$ is the standard unit vector in the $j^{th}$
direction.  The interactions between microstructures of the system can be readily handled
in the same manner as in the molecular dynamics simulation.  This is done by 
shifting the location of any microstructure of a periodic image involved in
an interaction, see Figure~\ref{figure_meshLeesEdwardsI}.

In practice, these boundary conditions present significant challenges for 
the numerical discretization of the fluid equations.  The conditions 
introduce both a jump discontinuity at periodic boundaries and a 
shift.  For uniform discretizations typically used for the unit cell, 
this results in significant misalignments of the nodes at the domain 
boundaries and a degradation in accuracy, see Figure~\ref{figure_meshLeesEdwardsI}.
When incorporating stochastic driving fields to account for thermal fluctuations 
these issues are further compounded.

To address these issues, we reformulate the momentum equations
in terms of a reference frame that is more naturally suited to the 
deformation of the material.  This is achieved by letting 
$\mathbf{w}(\mathbf{q},t) = \mathbf{u}(\phi(\mathbf{q},t),t)$,
where $\mathbf{q} = (q_1,q_2,q_3)$ parameterizes the deformed unit
cell.  The map from the moving coordinate frame to the fixed 
Eulerian coordinate frame is denoted by $\mathbf{x} = \phi(\mathbf{q})$
and given by $\phi(\mathbf{q},t) = (q_1 + q_3\dot{\gamma}t, q_2, q_3)$.
The SELM equations~\ref{equ_SELM_III_1}--~\ref{equ_SELM_III_3} in 
this reference frame are given by
\begin{eqnarray}
\label{equ_stokesShearFrame_momentum}
{\rho\frac{d\mathbf{w}^{(d)}}{dt}} & = & 
\mu\tilde{\nabla}^2{\mb{w}}
- \nabla{p} 
+ \OpSFCouple[-\nabla_{\mb{X}} \Phi] + \left(\nabla_{\mathbf{X}}\cdot\OpSFCouple\right) k_B{T}
+ \mathbf{J} + \mb{G}_\subtxt{thm} \\
\label{equ_stokesShearFrame_continuity}
\nabla
\cdot\mathbf{w}
 & = &
\mathbf{e}_z^T
\hspace{0.1cm}
\nabla\mathbf{w}
\hspace{0.1cm}
\mathbf{e}_x 
\dot{\gamma}t  
+
\mathbf{K} \\
\frac{d\mathbf{X}}{dt}  & = & \OpFSCouple\mathbf{w} \\
\label{equ_stokesShearFrame_thermal}
\langle \mathbf{G}_\subtxt{thm}(s)\mathbf{G}_\subtxt{thm}^{\RegAdjoint}(t) \rangle & = & 
-\left(2k_B{T}\right)\mu\tilde{\nabla}^2\hspace{0.06cm}\delta(t - s).
\end{eqnarray}
The Laplacian of the velocity field under the change of variable is given by
\begin{eqnarray}
\left[\tilde{\nabla}^2{\mb{w}}\right]^{(d)} & = & 
\left[
\mathbf{e}_d
-
\delta_{d,3}
\dot{\gamma}t
\mathbf{e}_x
\right]^T
\nabla^2 \mathbf{w}^{(d)}
\left[
\mathbf{e}_d
-
\delta_{d,3}
\dot{\gamma}t
\mathbf{e}_x
\right].
\end{eqnarray}
In this reference frame the misalignment arising from the shear boundary 
conditions is removed.  The boundary conditions for the reformulation
become formally periodic boundary conditions
\begin{eqnarray}
\mathbf{w}(q_1,q_2,L,t) = \mathbf{w}(q_1,q_2,0,t).
\end{eqnarray}
The jump discontinuities arising from the boundary condition~\ref{equ_StokesJumpBndCond} 
still remain but in the reformulation are now taken into account by introducing the terms 
$\mathbf{J},\mathbf{K}$.  The term $\mb{G}_\subtxt{thm}$ denotes the stochastic 
driving field accounting for thermal fluctuations in the moving frame of reference.

The
$\mathbf{q} = (q_1,q_2,q_3)$ parameterizes the deformed unit cell,
$\delta_{k,\ell}$ denotes the Kronecker $\delta$-function, 
$\dot{\gamma}$ denotes the rate of the shear deformation, and $\mathbf{e}_i$ 
the standard basis vector in the $i^{th}$ direction with $i \in \{x,y,z\}$.
In the notation the parenthesized superscript denotes a vector component.
We also use the notational conventions 
\begin{eqnarray}
\label{equ_components_Laplace_w}
\label{equ_components_grad_w}
\left[\nabla \mathbf{w}\right]_{j}^{(d)} 
=
\frac{\partial \mathbf{w}^{(d)}}{\partial q_j},
\hspace{1cm}
\left[\nabla^2 \mathbf{w}\right]_{i,j}^{(d)}
=
\frac{\partial^2 \mathbf{w}^{(d)}}{\partial q_i \partial q_j}.
\end{eqnarray}

\section{Computational Methodology}

\label{sec_comp_method}
To use this approach in practice requires 
the development of numerical methods to approximate 
the stochastic differential equations.  A variety of methods could be used, such
as Finite Difference Methods, Spectral Methods, or 
Finite Element Methods~\cite{Gottlieb1993,Strikwerda2004,Strang2008}.  
We present here discretizations based on Finite Difference Methods.
A challenge for discretizations is that the solutions to stochastic equations
are irregular and approximation errors may disrupt the statistical mechanics 
structure of the system.  To obtain physically relevant results, we shall design our 
numerical methods by developing a finite dimensional dynamical system that 
approximates the continuum mechanics while adhering to principles from statistical 
mechanics.  To introduce the stochastic driving fields in our numerical methods, we develop
discretizations that take into account the difference between the dissipative properties
of the continuum operators and the discrete operators so that our methods 
satisfy fluctuation-dissipation balance~\cite{Atzberger2007a, AtzbergerSELMRed2013, DeFabritiis2007a, Ottinger1997a, VazquezQuesada2012}.  
These conditions in the abscence of shear are closely related to the requirement that 
the Gibbs-Boltzmann ensemble be invariant with detailed-balance~\cite{Atzberger2007a, AtzbergerSELMRed2013}.
To perform simulations in practice we also develop efficient methods for generating the stochastic 
driving fields with the required covariance structure.
  
The stochastic differential equations of SELM can also exhibit dynamics over a significant range of time-scales
resulting in numerical stiffness~\cite{Atzberger2007a, AtzbergerSELMRed2013}.  To facilitate the development of efficient numerical 
methods, we consider two distinct physical regimes.  In the first we explicitly 
resolve the fluctuations and relaxation of the hydrodynamics.  We refer 
to this as the Fluctuating Hydrodynamics Regime.  In the second
we treat the solvent fluid as having relaxed to a quasi-steady-state
with respect the instantaneous configuration of the microstructures.
We refer to this as the Overdamped Regime.

\subsection{Numerical Methods for the Fluctuating Hydrodynamics Regime}
\label{sec_comp_method_fld_hyd}

\subsubsection{Semi-discretization} 
\label{sec_semidiscr_fld_hyd}
To approximate the stochastic differential equations
and to discretize the stochastic driving fields, we 
first consider a semi-discretization of the 
equations~\ref{equ_stokesShearFrame_momentum}--~\ref{equ_stokesShearFrame_thermal}.
This is given by 
\begin{eqnarray}
\label{equ_semiDiscr}
\rho \frac{d \mb{w}}{dt} & = & L(t) \mb{w} + \lambda + \OpSFCouple[-\nabla_{\mb{X}} \Phi] + \left(\nabla_{\mathbf{X}}\cdot\OpSFCouple\right) k_B{T}  + \mb{J} + \mb{h}_\subtxt{thm} \\
\label{equ_incomp_discr}
S(t)\cdot \mathbf{w} & = &
\mathbf{K} \\
\label{equ_struct_discr}
\frac{d \mb{X}}{dt}      & = & \OpFSCouple \mb{w}.
\end{eqnarray}
We use for the discretized operators 
\begin{eqnarray}
\label{equ_discr_S_t}
S(t)\cdot \mb{w}          & = & D\cdot \mb{w} + \mathbf{e}_z^T\hspace{0.1cm}G\mathbf{w}\hspace{0.1cm}\mathbf{e}_x \dot{\gamma}t   \\
\label{equ_discr_L_t}
L(t)\mb{w}                & = & \mu \left[\mathbf{e}_d - \delta_{d,3}\dot{\gamma}t\mathbf{e}_x\right]^T A\mb{w} \left[\mathbf{e}_d - \delta_{d,3} \dot{\gamma}t \mathbf{e}_x \right]
\end{eqnarray}
where
\begin{eqnarray}
D\cdot \mb{w}             & = & \sum_{d=1}^3 \frac{\mathbf{w}^{(d)}(\mathbf{q} + \mathbf{e}_d) - \mathbf{w}^{(d)}(\mathbf{q} - \mathbf{e}_d)}{2\Delta{x}}\\
\left[G\mb{w}\right]_{ij} & = & \frac{\mathbf{w}^{(i)}(\mathbf{q} + \mathbf{e}_j) - \mathbf{w}^{(i)}(\mathbf{q} - \mathbf{e}_j)}{2\Delta{x}}
\end{eqnarray}
and
\begin{eqnarray}
\left[A\mb{w}\right]_{ii} & = & \frac{\mathbf{w}^{(i)}(\mathbf{q} + \mathbf{e}_i) - 2\mathbf{w}^{(i)}(\mathbf{q}) + \mathbf{w}^{(i)}(\mathbf{q} - \mathbf{e}_i)}{\Delta{x}^2} \\
\left[A\mb{w}\right]_{ij} & = & \frac{\mathbf{w}^{(d)}(\mathbf{q} + \mathbf{e}_i + \mathbf{e}_j) - \mathbf{w}^{(d)}(\mathbf{q} - \mathbf{e}_i + \mathbf{e}_j)}{4\Delta{x}^2} \\
\nonumber
                          & - & \frac{\mathbf{w}^{(d)}(\mathbf{q} + \mathbf{e}_i - \mathbf{e}_j) - \mathbf{w}^{(d)}(\mathbf{q} - \mathbf{e}_i - \mathbf{e}_j)}{4\Delta{x}^2}, 
                                \mbox{ $i \not= j$}.
\end{eqnarray}
For the semi-discretized system we consider the energy 
\begin{eqnarray}
\label{equ_energy_discr}
E[\mb{w},\mb{X}] = \frac{\rho}{2} \sum_{\mb{q}} |\mb{w}(\mb{q})|^2 \Delta{x}_{\mb{q}}^3 + \Eng[\mb{X}].
\end{eqnarray}
The first term is the total kinetic energy of the system.  The second term is the potential energy
of the microstructures.  

For the discretized equations and energy, the $\mb{w}$ denotes the velocity field of the 
fluid on a uniform periodic lattice in the coordinates $\mb{q}$ with 
$\mb{w} \in \mathbb{R}^{3N}$.  The $N$ denotes the number of lattice sites.
The $\mb{X}$ denotes a finite number of microstructure degrees of 
freedom with $\mb{X} \in \mathbb{R}^M$.  As a consequence of the 
coordinate frame moving with the deformation of the unit cell,
the discretized operators now have a direct dependence on 
time.  The $\lambda$ denotes a Lagrange multiplier used to
impose the incompressibility condition~\ref{equ_incomp_discr}
and will be discussed in more detail below.

To obtain the source terms $\mathbf{J}$, $\mathbf{K}$ 
for the discretized equations, we use the discretization stencils
of the operators given in equations~\ref{equ_semiDiscr}--~\ref{equ_incomp_discr}.
When the stencils weights are applied at the boundaries 
of the unit cell, the values at lattice sites crossing 
the boundary would use the modified image value 
$\mathbf{w}_{\mathbf{m}} \pm \dot{\gamma}L$ under the 
boundary conditions~\ref{equ_StokesJumpBndCond}.
The use in the stencils of this modified lattice site value 
can be avoided by separating the contributions coming
from the jump part of the boundary condition from the usual
lattice site value of a periodic image.  These contributions 
are given by the stencil weights multiplied by 
$\pm \dot{\gamma}L$ for any term crossing the boundary.
When these are collected over all boundary mesh sites
and terms on the right-hand side involving $\mb{w}$, 
we obtain the source terms $\mathbf{J}$, $\mathbf{K}$.  

The incompressibility constraint for the solvent fluid is approximated 
in practice using the projection of a vector $\mathbf{v}^*$ to the sub-space 
$\{\mathbf{v} \in \mathbb{R}^{3N} \hspace{0.025cm} | \hspace{0.125cm} S(t)\cdot\mathbf{v} = 0\}$. 
We denote this projection operation by 
\begin{eqnarray}
\mathbf{v} = \wp(t) \mathbf{v}^*.
\end{eqnarray}
The discretized incompressibility constraint is imposed by using the Lagrange multiplier 
\begin{eqnarray}
\lambda = -(\mathcal{I} - \wp(t))
\left[L(t) \mb{w} + \OpSFCouple[-\nabla_{\mb{X}} \Phi] + 
\left(\nabla_{\mathbf{X}}\cdot\OpSFCouple\right) k_B{T}  + \mb{J} + \mb{h}_\subtxt{thm}\right]. 
\end{eqnarray}
We remark that the incompressibility constraint is imposed exactly 
provided that $\mb{K}$ is independent of time.  In practice, $\mb{K}$
is expected to have some dependence on time so that this approach 
results in an approximation in imposing the incompressibility 
constraint~\ref{equ_incomp_discr}.

An important feature of the discretization for the 
SELM equations and incompressibility constraint 
is that the resulting operators are cyclic.  This 
allows for Fast Fourier Transforms (FFTs) to be used
in evaluating the action of the operators and in
computing inverses.  As a consequence, the projection
operator can be computed efficiently with 
only $O(N\log(N))$ computational steps. 

To obtain appropriate behaviors for the thermal fluctuations, it is important
to develop stochastic driving fields which are tailored to the specific semi-discretization
used.  Another important issue is to develop methods for efficient generation of
the stochastic fields.  Once these issues are resolved, which is the subject of the 
next few sections, the semi-discretized equations can be integrated in time using traditional methods 
for stochastic differential equations, such as the Euler-Maruyama Method or a Stochastic 
Runge-Kutta Method~\cite{Platen1992}.  More sophisticated integrators in time can also 
be developed to cope with possible sources of stiffness~\cite{Atzberger2007a}.

\subsubsection{Thermal Fluctuations}
\label{sec_thermal_fld_hyd}
To account for thermal fluctuations, we introduce 
into the discretized equations a stochastic
driving field.  Given the highly irregular nature of the stochastic 
driving fields in the undiscretized equations~\ref{equ_SELM_III_1}--~\ref{equ_SELM_III_3}, 
formulating appropriate terms for the discretized 
equations must be done carefully. To obtain 
results consistent with statistical mechanics, we 
consider the relationship between the choice of 
stochastic driving field and the equilibrium 
fluctuations expected for the system.  To simplify 
the discussion, we initially consider only the case when 
$\Phi = 0$ and neglect the $\mb{X}$ degrees of freedom.
We then discuss how the results obtained 
apply to the more general case.

The statistical mechanics of the system requires equilibrium fluctuations 
which follow the Gibbs-Boltzmann distribution 
\begin{eqnarray}
\Psi(\mb{w}, \mb{X}) = \frac{1}{Z}\exp\left[-E[\mb{w},\mb{X}]/k_B{T}\right].
\end{eqnarray}
The $Z$ is the normalization constant ensuring the probability integrates to
one.  The $k_B$ is Boltzmann's constant and $T$ is the temperature~\cite{Reichl1998}.  
By considering the energy associated with
the discretized system given in equation~\ref{equ_energy_discr}, we see that fluctuations 
of $\mb{w}$ are Gaussian under the Gibbs-Boltzmann distribution.
This specific form of the energy along with the incompressibility constraint requires 
equilibrium fluctuations that have mean zero and covariance given by
\begin{eqnarray}
C = \langle \mb{w}\mb{w}^T \rangle = \frac{2}{3}\frac{k_B{T}}{\rho\Delta{x}^3}  \mathcal{I}.
\end{eqnarray}
The factor of $2/3$ arises from the incompressibility constraint.
The stochastic driving field $\mb{h}_\subtxt{thm}$ introduced into the 
discretized equations is assumed to be a Gaussian process with mean 
zero and $\delta$-correlation in time~\cite{Gardiner1985, Oksendal2000}.  
Such processes can be expressed formally as
\begin{eqnarray}
\mb{h}_\subtxt{thm} & = & Q(t) \frac{d \mb{B}(t)}{dt}.
\end{eqnarray}
The $Q(t)$ denotes a linear operator and 
$\mb{B}(t)$ denotes a standard Brownian motion on 
$\mathbb{R}^{3N}$, see~\cite{Oksendal2000}.  The covariance 
of this process is given by 
\begin{eqnarray}
G(s,t)  = 
\langle
\mb{h}_\subtxt{thm}(s)
\mb{h}_\subtxt{thm}(t)^T
\rangle
 = 
Q(s)Q(t)^T\delta(t - s). 
\end{eqnarray}
The discretized equations are linear in $\mb{w}$.
As a consequence, the covariance of the equilibrium 
fluctuations and the covariance of the stochastic 
driving field are related by 
\begin{eqnarray}
\label{equ_time_G}
G(s,t) = -2\wp(t)L(t)C \delta(t - s).
\end{eqnarray}
This relation can be interpreted as a variant of the 
fluctuation-dissipation principle.  We establish this
relationship for systems having time dependent 
dissipative operators in Appendix~\ref{appendix_fluct_dissip}.

This gives the the stochastic driving field $\mb{h}_\subtxt{thm}$ 
tailored to the moving coordinate frame and the  
semi-discretized equations~\ref{equ_semiDiscr}--~\ref{equ_struct_discr}.
By considering the Fokker-Planck
equations of the discretized system $(\mb{w},\mb{X})$, 
this choice can be shown to yield stochastic dynamics 
which have the Gibbs-Boltzmann 
distribution invariant, see~\cite{AtzbergerSELM2011}.
This shows the stochastic dynamics exhibit
fluctuations consistent with equilibrium 
statistical mechanics.  It should be mentioned,
evaluating the appropriateness of 
this choice for the stochastic driving field also can be 
investigated by considering other 
properties, such as the dynamic structure 
factor of the stochastic dynamics that 
could be used to make comparisons with 
the undiscretized equations or with
physical systems~\cite{Donev2009a,Mezei2003}.
Another important issue arising in practice 
is to develop computational methods for the efficient generation of 
the stochastic driving field.  
This is the subject of the next section.

\subsubsection{Generation of Stochastic Driving Fields}
\label{sec_gen_stoch_fld_hyd}

To account for thermal fluctuations, we must generate
each time step the Gaussian stochastic field with the covariance structure 
given by equation~\ref{equ_time_G}.  In general, generating 
a Gaussian variate $\mb{h}$ with a prescribed covariance $G$ 
is computationally expensive.  A common approach is 
to generate standard normal variates $\boldsymbol{\xi}$ 
having covariance $\langle \boldsymbol{\xi} \boldsymbol{\xi}^T \rangle = I$.
To obtain a correlated Gaussian a Cholesky factorization is often
used to obtain $QQ^T = G$ and the Gaussian is generated using 
$\mb{h} = Q \boldsymbol{\xi}$.  For $\mb{h} \in \mathbb{R}^N$,
the Cholesky factorization has a cost of 
$O(N^3)$ computational steps and the generation of each 
variate through the matrix-vector multiplication has 
a cost of $O(N^2)$ computational steps.  For the discretized
equations $N$ will typically be rather large making this
approach prohibitive.  

To generate the stochastic driving field more efficiently,
we make use of specific properties of the discretization
and FFTs.  These properties include that $\mb{w}$ is periodic in the 
moving coordinate frame and that the discretized 
operators $L(t)$, $C$, and $\wp(t)$ are block
diagonalizable in the Fourier basis (with blocks
of small size).  By working with the diagonalized form
of each of the operators $L(t)$, $C$, and $\wp(t)$, 
a square-root $Q(t)$ of the operator $G(t)$ can be 
found in Fourier space.  Given the sparse structure 
of $Q(t)$ in the Fourier space, the stochastic fields 
are generated using FFTs in $O(N\log(N))$ 
computational steps.

\subsection{Numerical Methods for the Overdamped Regime}
\label{sec_comp_method_overdamped}

For many physical systems of interest, there are significant
differences in the time scales associated with the hydrodynamic
relaxation of the solvent fluid and the time scales associated 
with the diffusion of the microstructures an appreciable distance.  
For the fluctuating hydrodynamics regime this can result in 
significant stiffness in the stochastic differential equations.   

For investigations of complex fluids and soft materials in which 
the relaxation of the hydrodynamics is not of primary interest, 
it is useful to introduce a reduced description removing this 
source of stiffness.  In the limit of fluid dynamics which rapidly 
equilibrate given the instantaneous configuration of the 
microstructures we have the reduced equations, 
\begin{eqnarray}
\label{equ_SELM_IV_1}
\frac{d\mathbf{X}}{dt}  & = & H_\subtxt{SELM}[-\nabla_{\mathbf{X}}\Phi(\mathbf{X})] 
+ (\nabla_{\mb{X}}\cdot H_\subtxt{SELM})k_B{T}
+ \mathbf{h}_\subtxt{thm} \\
\label{equ_SELM_IV_2}
H_\subtxt{SELM}         & = & \OpFSCouple(-\wp L)^{-1}\OpSFCouple \\
\label{equ_SELM_IV_3}
\langle \mathbf{h}_\subtxt{thm}(s)\mathbf{h}_\subtxt{thm}^{\RegAdjoint}(t) \rangle & = & \left(2k_B{T}\right)H_\subtxt{SELM}\hspace{0.06cm}\delta(t - s).
\end{eqnarray}
The $\wp$ denotes a projection operator imposing constraints, 
such as incompressibility.  The adjoint property   
$\OpSFCouple = \OpFSCouple^{\OpAdjoint}$
and symmetry of $\wp L$ yields an 
operator $H_\subtxt{SELM}$ which is symmetric.  
The semi-discretization of these equations is obtained by
using the discretized operators given in 
equations~\ref{equ_discr_S_t}--~\ref{equ_discr_L_t}.
The thermal fluctuations are determined by the principle
of detailed-balance and the requirement that the Gibbs-Boltzmann
distribution be invariant under the stochastic dynamics, see~\cite{AtzbergerSELM2011, AtzbergerSELMRed2013}.
The semi-discretized equations can be integrated 
in time using standard methods for stochastic differential equations~\cite{Platen1992}.  

We remark that when using the Immersed Boundary Method to coupled the fluid and structures the hydrodynamic 
coupling tensor closely resembles the Rotne-Prager-Yamakowa tensor~\cite{AtzbergerSELM2011}.   We remark that while
$H_\subtxt{SELM}$ accounts well for the far-field hydrodynamics, additional near-field corrections 
could also be added readily to the hydrodynamic coupling, such as terms to account for lubrication effects 
as is done in such methods as Stokesian-Brownian 
dynamics~\cite{Brady1988}.  However, for polymers and other soft microstructures the near-field interactions are less 
clear than at larger length-scales and ideally should be determined from more detailed molecular models and considerations.  
We also remark that the far-field hydrodynamic interactions pose the most computational challenge since they couple long-range 
the microstructures in a nearly all-to-all manner while the near-field interactions are local and can be handled for a small cluster around
each microstructure at much less computational expense.  Since these terms can be treated additively throughout, the methods 
we shall present for the long-range hydrodynamics can be extended readily to incorporate near-field interactions.   The central challenge for such methods in practice is the long-range hydrodynamics and the efficient generation of the stochastic driving field $\mathbf{h}_\subtxt{thm}$ 
with the required covariance structure given by equation~\ref{equ_SELM_IV_3}. 

\subsubsection{Generation of Stochastic Driving Fields}
\label{sec_gen_stoch_overdamped}

To use this description in practice requires efficient methods
for generating the stochastic driving field with the covariance
given in equation~\ref{equ_SELM_IV_3}.  For this purpose we 
express the covariance of the stochastic driving field as
\begin{eqnarray}
\label{equ_G_regime_IV}
G = (2k_B{T})H_\subtxt{SELM} = (2k_B{T}) \left( \OpFSCoupleDiscr  \wp (-\OpFldDisspDiscr)^{-1} \wp^{\RegAdjoint} \OpFSCoupleDiscr^{\RegAdjoint} \right).
\end{eqnarray}
This makes use of $\OpSFCoupleDiscr = \OpFSCoupleDiscr^{\RegAdjoint}$ and properties of the 
specific discretized operators $\OpFldDisspDiscr$ and $\wp$.
In particular, commutativity $\wp \OpFldDisspDiscr = \OpFldDisspDiscr \wp$ and the projection 
operator properties $\wp^2 = \wp$, $\wp = \wp^{\RegAdjoint}$.  Let $U$ be a factor so that 
$UU^{\RegAdjoint} = -\OpFldDisspDiscr^{-1}$.  Using this factor we can express the covariance as 
\begin{eqnarray}
\label{equ_G_spec_factor_IV}
G = \left( \sqrt{2k_B{T}} \OpFSCoupleDiscr\wp U \right) \left( \sqrt{2k_B{T}} \OpFSCoupleDiscr\wp U \right)^{\RegAdjoint}.
\end{eqnarray}
From this expression a matrix square-root of $G$ is readily obtained,
$Q = \sqrt{2k_B{T}} \OpFSCoupleDiscr \wp U$.

We remark this is different than 
the Cholesky factor obtained for $G$ which is 
required to be lower triangular~\cite{Trefethen1997, Strang1988}.
Obtaining such a factor by Cholesky factorization would cost 
$O(M^3)$, where $M$ is the number of structure degrees of freedom.  
For the current discretization considered, the operators $L$ and 
$\wp$ are block diagonalizable in Fourier space (with small blocks).  This 
has the consequence that the action of the operators $U$ and $\wp$ 
can be computed using FFTs with a cost of $O(N \log(N))$.  The $N$ is the 
number of lattice sites used to discretize $L$.  The stochastic driving field 
is computed from $\mathbf{h} = Q\boldsymbol{\xi}$.  This allows for the 
stochastic driving field to be generated in $O(N\log(N) + M)$ computational 
steps, assuming the action $\OpFSCoupleDiscr$ can be compute in $O(M)$ steps.  
This is in contrast to using the often non-sparse matrix arising from 
Cholesky factorization which generates the stochastic field with a cost
of $O(M^2)$.  Other methods based on splittings or 
multigrid can also be utilized to efficiently generate stochastic fields
with this required covariance structure or for discretizations on 
multilevel adaptive meshes, see~\cite{AtzbergerSELM2011,Atzberger2010a}.

\subsection{Operators for Coupling the Microstructures and Solvent Fluid}
\label{sec_operators}
Many different operators could be used to couple the 
microstructure and solvent dynamics~\cite{AtzbergerSELM2011}.  
We will take an approach similar to the Stochastic 
Immersed Boundary Method~\cite{Atzberger2007a,Peskin2002}
and use the following specific form for the operators that 
couple the microstructures and solvent fluid
\begin{eqnarray}
\label{equ_IB_coupling}
\left[\OpFSCouple \mathbf{u}\right](\mb{s})
& = & 
\int_{\Omega} \eta(\mathbf{y} - \mathbf{X}(\mb{s}))
\mathbf{u}(\mb{y}) d\mb{y} \\
\left[\OpSFCouple \mathbf{F}\right](\mb{y})
& = & 
\int_{\mathcal{S}} \mathbf{F}(\mb{s}) 
\eta(\mathbf{y} - \mathbf{X}(\mb{s})) d\mb{s}.
\end{eqnarray}
The $\mb{F}$ denotes
the force acting on the microstructures,
which is typically given by $\mb{F} = -\nabla_{\mb{X}} \Phi$.
The kernel function $\eta$ is used to smooth
the irregular velocity field and determines
an effective hydrodynamic radius for the 
microstructures, see~\cite{AtzbergerSELM2011,Atzberger2007a}
and Appendix~\ref{appendix_delta_func}.  
It can be shown these operators satisfy 
the adjoint condition given by 
equation~\ref{equ_adjoint_cond}.  
This pair of operators has been successfully used in the past 
and extensive validation studies have been conducted to characterize
how these operators represent hydrodynamic 
coupling~\cite{Atzberger2007a, Atzberger2007c, Atzberger2008, AtzbergerSELM2011, Kramer2008, Bringley2008}.

To couple the semi-discretized description
of the solvent fluid and microstructures
we use the discretized operators 
\begin{eqnarray}
\label{equ_IB_coupling_discr}
\left[\OpFSCouple \mathbf{u}\right]^{[j]}
& = & 
\sum_{\mathbf{m}} \eta(\mathbf{y}_{\mathbf{m}} - \mathbf{X}^{[j]})
\mathbf{u}_{\mathbf{m}} \Delta{y}_{\mb{m}}^d \\
\left[\OpSFCouple \mathbf{F}\right]_{\mb{m}}
& = & 
\sum_{j = 1}^{M} \mathbf{F}^{[j]}(\mathbf{X}) 
\eta(\mathbf{y}_{\mathbf{m}} - \mathbf{X}^{[j]}).
\end{eqnarray}
It can be shown these operators satisfy an 
adjoint condition analogous to equation~\ref{equ_adjoint_cond}
for the semi-discretized equations~\ref{equ_semiDiscr}--~\ref{equ_struct_discr}.  
More general discretized operators can also be used, 
see~\cite{AtzbergerSELM2011}.

\section{Estimating Macroscopic Stresses}
\label{sec_stress_estimator}

\begin{figure}[t]
\centering
\includegraphics[width=5in]{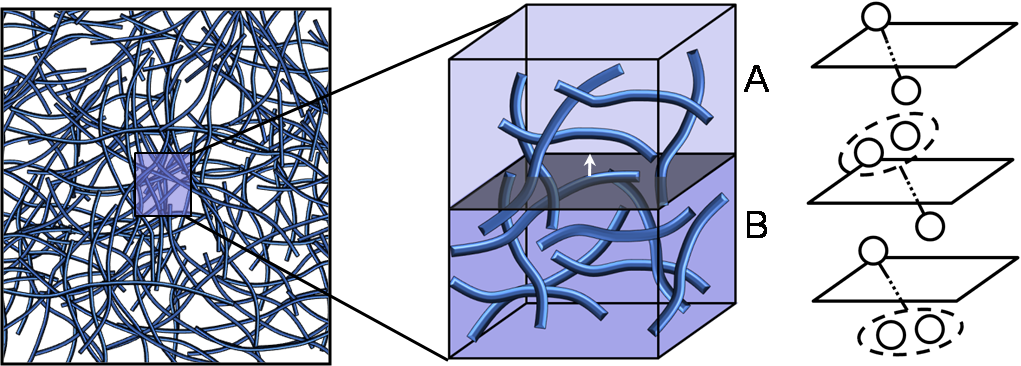}
\caption[Macroscopic Stress Estimator]
{Estimating Macroscopic Stresses.  The components of the
stress tensor are estimated from microscopic interactions
by considering cut-planes through the sample.  The 
cut-plane divides the sample into two bodies labeled 
$A$ and $B$, shown in the middle.  The effective stress 
is given by the forces exerted by the material of body 
$A$ on material of body $B$ divided by the cut-plane area.  
For $n$-body interactions, 
contributions to the stress can arise in different
ways.  On the far right is shown for a given cut-plane 
the cases when two-body and three-body interactions make a 
contribution to the stress. 
}
\label{figure_stressEstimator}
\end{figure}

An important issue in the study of complex fluids and soft materials
is to relate microscopic features of the material to macroscopic 
properties.  For this purpose, we develop 
estimators for an effective macroscopic stress tensor for the 
material.  The stress tensor is estimated from the forces 
acting on the microstructures in a manner similar to that 
used in the Irving-Kirkwood-Kramer 
formulas~\cite{Irving1950,Doi1986,Bird1987Vol1,Bird1987Vol2}.
The contributions from the solvent fluid are assumed 
to be Newtonian throughout.

The stress tensor is estimated by considering 
cut-planes which divide the unit cell.  A
component of the stress tensor is estimated 
by considering the forces exerted by the material 
which lies above the cut-plane on the material which 
lies below the cut-plane.  The totality of these 
forces is then divided by the area of the cut-plane.  
We average these estimates over all possible cut-planes having a given normal
to avoid sensitive dependence on the microstructure 
configuration, the $n$-body interactions, and the cut-plane 
location, see Figure~\ref{figure_stressEstimator}.

To estimate the stress tensor, it is convenient to consider 
separately each of the different types of $n$-body interactions 
which occur between the microstructures, such as two-body
bonding interactions or three-body bond-angle interactions.
To estimate the contributions to the components of the 
stress tensor arising from a particular $n$-body interaction, we use
\begin{eqnarray}
\label{equ_init_stress_estimator}
\sigma_{\ell,z}^{(n)} = \frac{1}{L} \left \langle \int_a^b \Theta_{\ell,z}^{(n)}(\zeta) d\zeta \right \rangle.
\end{eqnarray}
The $L = b - a$ is the length of the unit-cell domain in the $z$-direction
and
$\langle\cdot\rangle$ denotes averaging over the ensemble.
The $\Theta_{\ell,z}^{(n)}$ denotes the effective 
stress arising from the $n$-body interactions associated 
with a given stress plane and is defined by
\begin{eqnarray}
\label{equ_micro_stress_estimator}
\Theta_{\ell,z}^{(n)}(\zeta) 
& = & 
\frac{1}{A} 
\sum_{\mathbf{q} \in \mathcal{Q}_n}
\sum_{k = 1}^{n - 1}
\sum_{j = 1}^{k}
\mathbf{f}_{\mathbf{q}, j}^{(\ell)}
\prod_{j = 1}^{k}
\mathcal{H}(\zeta - \mathbf{x}_{q_j}^{(z)})
\prod_{j = k + 1}^{n}
\mathcal{H}(\mathbf{x}_{q_j}^{(z)} - \zeta).
\end{eqnarray}
The $\mathcal{Q}_{n}$ is the set of $n$-tuple 
indices $\mathbf{q} = (q_1,\ldots,q_n)$
describing the $n$-body interactions of the system,
$\mathbf{f}_{\mathbf{q}, j}$ denotes the force
acting on the $j^{th}$ particle of the interaction,
and $\mathbf{x}_{q_j}$ denotes the $j^{th}$ 
particle involved in the interaction.  As a matter 
of convention in the indexing $\mathbf{q}$,
we require that
$i \leq j$ implies $\mathbf{x}_{q_i}^{(z)} 
\leq \mathbf{x}_{q_j}^{(z)}$.  This 
expression corresponds to a sum over all 
the forces exerted by particles of the material
above the cross-section at $\zeta = z$ on the 
particles of the material below.  Each term
of the summation over $k = 1, \ldots, n - 1$
corresponds to a specific number of particles 
of the $n$-body interaction lying below the 
cross-section at $\zeta = z$, see Figure~\ref{figure_stressEstimator}.

This expression for estimating the stress tensor can be simplified 
by using the following identity
\begin{eqnarray}
\label{equ_identity_stress}
\int_a^b \Pi_{j = 1}^{k}
\mathcal{H}(\zeta - \mathbf{x}_{q_j}^{(z)})
\cdot
\Pi_{j = k + 1}^{n}
\mathcal{H}(\mathbf{x}_{q_j}^{(z)} - \zeta) 
d\zeta = \mathbf{x}_{q_{k + 1}}^{*,(z)} - \mathbf{x}_{q_k}^{*,(z)}
\end{eqnarray}
where 
\begin{eqnarray}
\mathbf{x}_{q_j}^{*,(z)} = 
\left\{
\begin{array}{ll}
b,                                & \mbox{if $\mathbf{x}_{q_j}^{(z)} \geq b$} \\
\mathbf{x}_{q_j}^{(z)},           & \mbox{if $a \leq \mathbf{x}_{q_j}^{(z)} \leq b$} \\
a,                                & \mbox{if $\mathbf{x}_{q_j}^{(z)} \leq a$}. \\
\end{array}
\right.
\end{eqnarray}
By integrating equation~\ref{equ_micro_stress_estimator} and using the identity 
given in equation~\ref{equ_identity_stress}, we obtain
\begin{eqnarray}
\label{equ_identity_stress2}
\int_a^b
\Theta_{(\ell),z}^{(n)}(\zeta) 
d\zeta
= \frac{1}{A} 
\sum_{\mathbf{q} \in \mathcal{Q}_n}
\sum_{k = 1}^{n - 1}
\sum_{j = 1}^{k}
\mathbf{f}_{\mathbf{q}, j}^{(\ell)}
\cdot
\left(
\mathbf{x}_{q_{k + 1}}^{*,(z)} - \mathbf{x}_{q_k}^{*,(z)}
\right).
\end{eqnarray}
This can be further simplified by switching the order of summation of $j$ and $k$
and using the telescoping property of the summation over $k$.  This gives
the following estimate for the $n$-body contributions to the stress tensor
\begin{eqnarray}
\label{equ_stress_estimator}
\sigma_{\ell,z}^{(n)} = \frac{1}{AL} 
\sum_{\mathbf{q} \in \mathcal{Q}_n}
\sum_{j = 1}^{n - 1}
\left \langle 
\mathbf{f}_{\mathbf{q}, j}^{(\ell)}
\cdot
\left(
\mathbf{x}_{q_n}^{*,(z)} - \mathbf{x}_{q_j}^{*,(z)}
\right)
\right \rangle.
\end{eqnarray}
This is obtained by using equation~\ref{equ_init_stress_estimator}
and equation~\ref{equ_identity_stress2}.

To obtain the effective macroscopic stress tensor
we sum over all $n$-body contributions to obtain
\begin{eqnarray}
\sigma_{\ell,z} = \sum_n \sigma_{\ell,z}^{(n)}.
\end{eqnarray}
This effective macroscopic stress tensor will be used to link
the microscopic simulations to macroscopic material properties.

\section{Applications}
\label{sec_applications}

To demonstrate how the computational methods can be used to study the 
rheological behaviors of complex fluids and soft materials, 
we present a few specific applications.  
We investigate the shear thinning of a polymeric fluid
in Section~\ref{sec_appl_shear_thinning}.  We study the complex moduli 
for the oscillatory responses of a polymerized lipid 
vesicle in Section~\ref{sec_appl_osc_responses}.  We study the 
aging of the shear viscosity of a gel-like material in 
Section~\ref{sec_appl_aging_shear_viscosity}.

\subsection{Application I: Shearing Thinning of a Polymeric Fluid}

\label{sec_appl_shear_thinning}

\begin{figure}[t]
\centering
\includegraphics[width=5in]{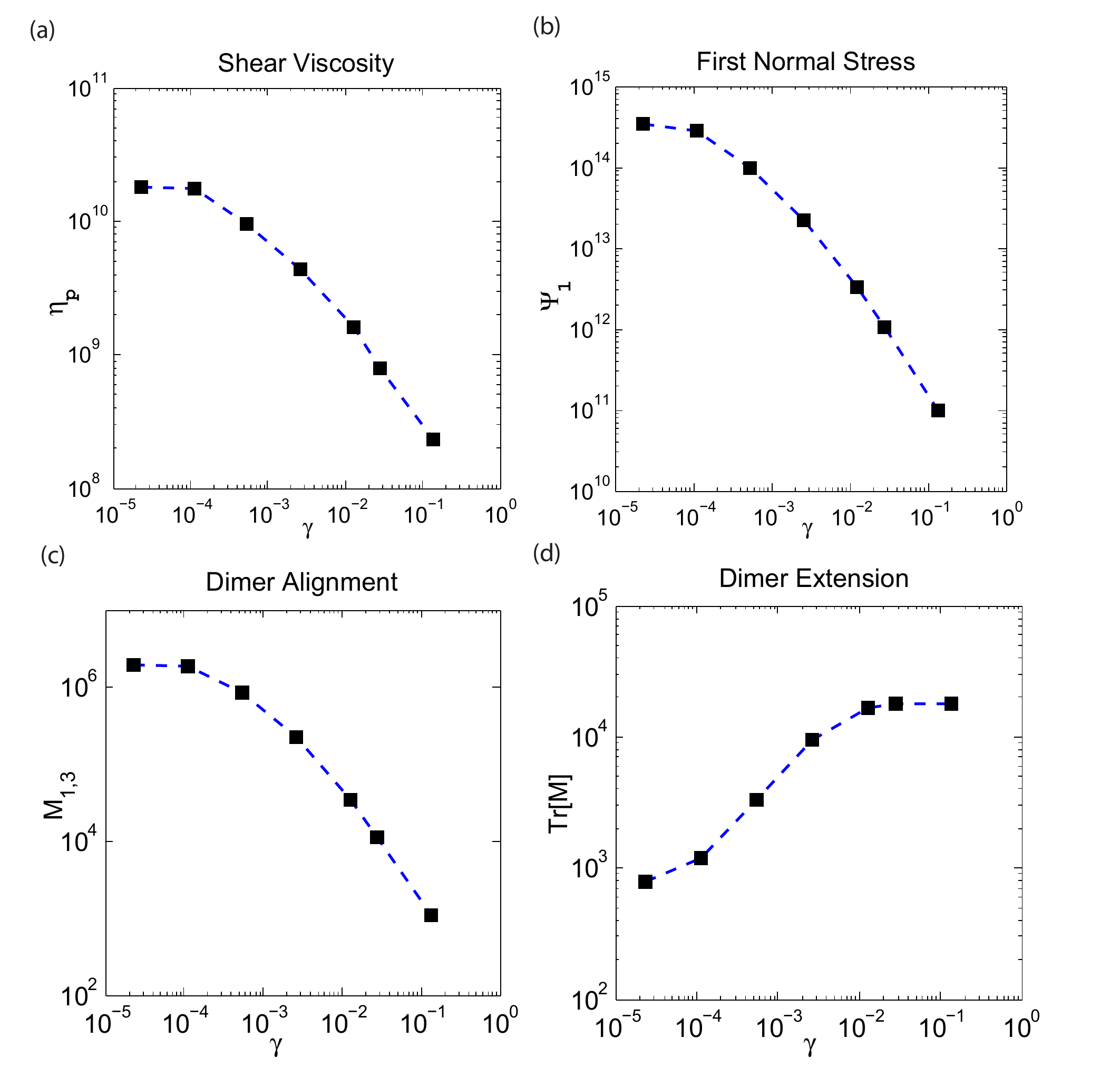}
\caption[Polymeric Fluid Simulation Results]
{Polymeric Fluid Shear Response : Simulation Results.  (a--b) The shear 
viscosity and first normal stresses of the polymeric fluid in response
to shear at the rate $\dot{\gamma}$.
To quantitate the alignment and extension of the polymers, the 
orientation matrix is considered given by 
$M = \langle \mathbf{z} \mathbf{z}^T\rangle$, 
with $\mb{z} = \mb{X}^{(2)} - \mb{X}^{(1)}$.  (c) The component 
$M_{1,3}$ gives a measure of the alignment of the polymers 
with the direction of shear.  (d) The trace of $M$ gives 
a measure of the extension of the polymers.
}
\label{figure_polymeric_fluid_results}
\end{figure}

As a demonstration of the proposed computational methodology we  
consider a fluid with microstructures consisting of elastic polymers.
The polymers are modeled as elastic dimers
which have the FENE potential energy~\cite{Bird1987Vol2} 
\begin{eqnarray}
\label{equ_V_r}
\phi(r) = \frac{1}{2} K r_0^2 \log\left(1 - \left(\frac{r}{r_0}\right)^2\right).
\end{eqnarray}
The $K$ denotes the polymer stiffness, $r$ denotes the length of extension of the 
dimer, and $r_0$ denotes the maximum permitted extension length.  
The configuration of a dimer will be represented using 
two degrees of freedom $\mathbf{X}_k^{(1)}$, $\mathbf{X}_k^{(2)}$ with
the potential energy $\Phi(\mathbf{X}) = \sum_k \phi(|\mathbf{X}_k^{(2)} - \mathbf{X}_k^{(1)}|)$.
The $\mathbf{X}$ denotes the composite vector over all dimers.

To study the rheology of the polymeric fluid, we consider 
the shear viscosity $\eta_p$ and the first normal stress coefficient $\Psi_1$.  
These are defined as~\cite{Bird1987Vol1,Bird1987Vol2}
\begin{eqnarray}
\label{equ_eta_p_def_FENE}
\eta_p & = & {\sigma_p^{(s,v)}}/{\dot{\gamma}} \\
\Psi_1 & = & ({\sigma_p^{(s,s)} - \sigma^{(v,v)}})/{\dot{\gamma}^2}.
\end{eqnarray}
The $\dot{\gamma}$ is the rate of shear.
In the notation, the superscript $(s,v)$ indicates the 
tensor component with the index $s$ corresponding to the 
direction of the shear gradient and the index $v$ corresponding to 
the direction of the velocity induced by the shear.  The contributions of the 
solvent fluid to the shear viscosity and normal stresses can be 
considered separately~\cite{Bird1987Vol2}.  The solvent fluid
is assumed to be Newtonian throughout so we only report the 
contributions arising from the elastic dimers.

From the simulations, we find there is a strong dependence
on the rate of shear in the manifested shear viscosity
and normal stress of the polymeric fluid, see 
Figure~\ref{figure_polymeric_fluid_results}.  
This can be understood by considering
the interplay between the thermal fluctuations
and the shear stresses acting on the dimers.  
Since the dimers only resist stretching, 
they exert forces only in the direction of the dimer 
orientation.  As a consequence, contributions are 
made to the shear viscosity only when the dimer 
orientation has a non-negligible component in the 
z-direction, see equation~\ref{equ_eta_p_def_FENE}.

\begin{table}[h]
\centering
\begin{tabular}{|l|l|}
\hline
Parameter & Description \\
\hline
$N$                           & Number of mesh points in each direction.\\
$\Delta{x}$                   & Mesh spacing.                  \\
$L$                           & Domain size in each direction. \\
$T$                           & Temperature.                   \\
$k_B$                         & Boltzmann's constant. \\                              
$\mu$                         & Dynamic viscosity of the solvent fluid. \\                             
$\rho$                        & Mass density of the solvent fluid. \\                             
$K$                           & Bond stiffness. \\
$r_0$                         & Maximum permissible bond extension. \\
$\gamma_s$                    & Stokesian drag of a particle. \\
$\dot{\gamma}^0$              & Shear rate amplitude. \\
$\gamma^0$                    & Strain rate amplitude. \\
$a$                           & Effective radius of particle estimated via Stokes drag. \\
\hline
\end{tabular}
\caption[FENE Parameter Description]
{Description of the parameters used in simulations of the polymeric fluid.
\label{table_FENE_param_descr}}
\end{table}

\begin{table}[h]
\centering
\begin{tabular}{|l|l|}
\hline
Parameter & Value \\
\hline   
$N$                           & 36 \\
$\Delta{x}$                   & $11.25 \mbox{ nm}$ \\
$L$                           & $405 \mbox{ nm}$ \\
$T$                           & $300 \mbox{ K}$ \\
$k_B$                         & $8.3145 \times10^3    \mbox{ nm}^2\cdot\mbox{amu}\cdot\mbox{ns}^{-2}\cdot\mbox{K}^{-1}$ \\
$\mu$                         & $6.0221 \times 10^{5} \mbox{ amu}\cdot\mbox{cm}^{-1}\cdot\mbox{ns}^{-1}$ \\
$\rho$                        & $6.0221 \times 10^{2} \mbox{ amu}\cdot\mbox{nm}^{-3}$ \\
$K$                           & $8.9796 \times 10^3 \mbox{ amu}\cdot\mbox{ns}^{-2}$ \\
$r_0$                         & $200 \mbox{ nm}$ \\
$\gamma_s$                    & $1.7027 \times 10^8 \mbox{ amu}\cdot\mbox{ns}^{-1}$  \\
$a$                           & $15 \mbox{ nm}$ \\
\hline
\end{tabular}
\caption[FENE Parameter Values]
{Values of the parameters used in simulations of the polymeric fluid.
\label{table_FENE_value_descr}}
\end{table}

The thermal 
fluctuations and shear stresses play opposing 
roles with respect to the z-component.  
The thermal fluctuations act to randomize
the dimer orientation generating on average a
non-negligible z-component while the shear 
stresses act to align the dimers 
with the direction of shear and suppress the z-component.  
As the shear rate increases, this results
in an increase in the shear stresses and an increase 
in the degree of alignment of
the dimers.  This results in a decrease in the 
shear viscosity.  This can be quantitated in the 
simulations by considering for the dimers 
the orientation tensor 
$M = \langle \mb{z}\mb{z}^T \rangle$, where 
$\mb{z} = \mb{X}^{(2)} - \mb{X}^{(1)}$, see
Figure~\ref{figure_polymeric_fluid_results}.  
This highlights the important
roles that thermal fluctuations can play 
in material properties.  For this polymeric
fluid, if thermal fluctuations
were neglected, there would be no 
contributions to the shear viscosity by
the dimers since they would all eventually 
align with the direction of shear.

The observed decrease of the shear viscosity 
with an increase in the shear rate is a 
common phenomena observed for many 
complex fluids~\cite{Bird1987Vol2}.  This
behavior is referred to as ``shear-thinning``, 
see~\cite{Bird1987Vol2}.  These simulations 
give a proof-of-principle for how such 
phenomena can be studied for complex 
fluids using the presented computational 
methodology.  For the simulation parameters used in
our simulations see Tables~\ref{table_FENE_param_descr}
and Table~\ref{table_FENE_value_descr}.

\subsection{Application II: Complex Moduli for Oscillatory Responses of Polymerized Lipid Vesicles}

\label{sec_appl_osc_responses}

As a further demonstration of the computational methods,
we investigate the material properties of a fluid 
containing polymerized lipid vesicles.  We discuss how the 
methods can be used to study responses to an oscillatory
shear applied over a wide range of frequencies.

\begin{figure}[t]
\centering
\includegraphics[width=5in]{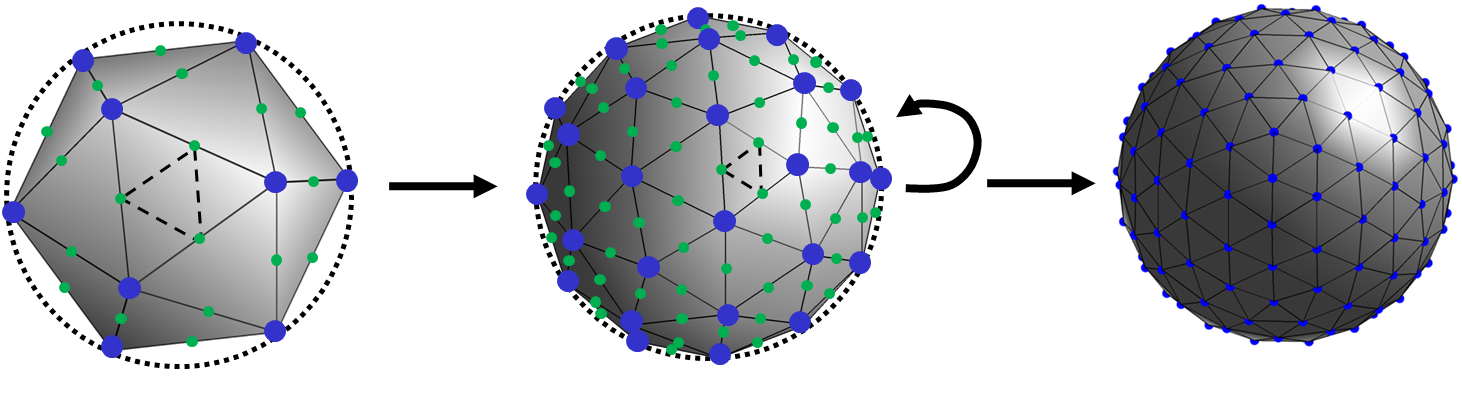}
\caption[Vesicle Mesh Construction]
{Vesicle Mesh Construction using a Recursive Refinement Method.  
The triangulated
mesh for a spherical vesicle is constructed by starting with 
the vertices and faces of a regular icosahedron, shown on the 
left.  The edges of the icosahedron are bisected and connected
to divide each triangular face into four smaller triangular 
faces.  The vertices located at the bisection points are 
projected radially outward to the surface of the sphere,
shown in the middle.  This refinement procedure is repeated 
recursively until a mesh of sufficient resolution is obtained.
The mesh obtained after two levels of recursive refinement
is shown on the far right.
}
\label{figure_vesicles_meshAlg}
\end{figure}

To account for the mechanics of a polymerized lipid vesicle,
we discretize the spherical surface using a triangular mesh.
The mechanics is modeled by the following interactions between the 
control points of the mesh 
\begin{eqnarray}
\label{equ_vesicle_energy_extend}
\phi_1(r,\ell) & = & 
\frac{1}{2}K_1\left(r - \ell\right)^2 \\
\label{equ_vesicle_energy_curvature}
\phi_2(\boldsymbol{\tau}_1,\boldsymbol{\tau}_2) & = & 
\frac{1}{2}K_1 \left|\boldsymbol{\tau}_{1} - \boldsymbol{\tau}_{2}\right|^2.
\end{eqnarray}
The $r$ denotes the displacement between two control points, $\ell$ denotes 
a preferred distance between control points, and $\boldsymbol{\tau}$ denotes
a normalized displacement vector (tangent vector) between two control points.
The $\phi_1$ energy accounts for the stretching of a bond between two control
points beyond its preferred extension.  The $\phi_2$ energy accounts for 
bending of the surface locally by penalizing the misalignment of tangent
vectors.  The total energy for a given configuration of the vesicle is
given by 
\begin{eqnarray}
\Phi[\mathbf{X}] & = & E_1[\mathbf{X}] + E_2[\mathbf{X}] \\
E_1[\mathbf{X}] & = & 
\sum_{(i,j) \in \mathcal{Q}_1} \phi_1(r_{ij},\ell_{ij}) \\
E_2[\mathbf{X}] & = & 
\sum_{(i,j,k) \in \mathcal{Q}_2} \phi_2(\boldsymbol{\tau}_{ij},\boldsymbol{\tau}_{jk}).
\end{eqnarray}
The $\mathbf{X}$ denotes the composite vector of control points.  The 
$j^{th}$ control point is denoted by $\mathbf{X}^{[j]}$.  The $\mathcal{Q}_1$
and $\mathcal{Q}_2$ are index sets defined by the topology of the triangulated 
mesh.

The first energy term $E_1$ accounts for stretching of the vesicle surface and is 
computed by summing over all local two body interactions $\mathcal{Q}_1$
defined by the topology of the triangulated mesh.  For the distance 
$r_{ij} = |\mathbf{X}^{[i]} - \mathbf{X}^{[j]}|$ between the two points 
having index $i$ and $j$, the energy $E_1$ penalizes deviations from
the preferred distance $\ell_{ij}$.
The preferred distances $\ell_{ij}$ are defined by the geometry 
of a spherical reference configuration for the vesicle.  To ensure the
two body interactions are represented by a unique index in $\mathcal{Q}_1$ 
we adopt the convention that $i < j$.

The second energy term $E_2$ accounts for curvature of the vesicle surface and 
is computed by summing over all local three body interactions $\mathcal{Q}_2$ 
defined by the topology of the triangulated mesh.  The energy penalizes the 
the misalignment of the tangent vectors 
$\boldsymbol{\tau}_{ij} = (\mathbf{X}^{[i]} - \mathbf{X}^{[j]})/r_{ij}$ and
$\boldsymbol{\tau}_{jk} = (\mathbf{X}^{[j]} - \mathbf{X}^{[k]})/r_{jk}$. 
In the set of indices in $\mathcal{Q}_2$ it is assumed
that the point with index $j$ is always adjacent to both 
$i$ and $k$.  To ensure the three body interactions are represented by a unique 
index in $\mathcal{Q}_2$ we adopt the convention that $i < k$.

To obtain a triangulated mesh which captures the shape of 
a vesicle having a spherical geometry we start with an 
icosahedral which is circumscribed by a sphere of a given 
radius.  We use the faces of the icosahedron as an initial
triangulated mesh.  To obtain a mesh which better approximates 
the sphere we bisect the three edges of each triangular face 
to obtain four sub-triangles.  The newly introduced vertices 
are projected radially outward to the surface of the sphere.  
The process is then repeated recursively to obtain further 
refinements of the mesh.  This yields a high quality mesh 
for spherical geometries.  We use a vesicle represented by a mesh
obtained using two levels of recursive refinement.  The recursive
generation procedure and the mesh used to represent a vesicle
is shown in Figure~\ref{figure_vesicles_meshAlg}.

\begin{figure}[t]
\centering
\includegraphics[width=5.in]{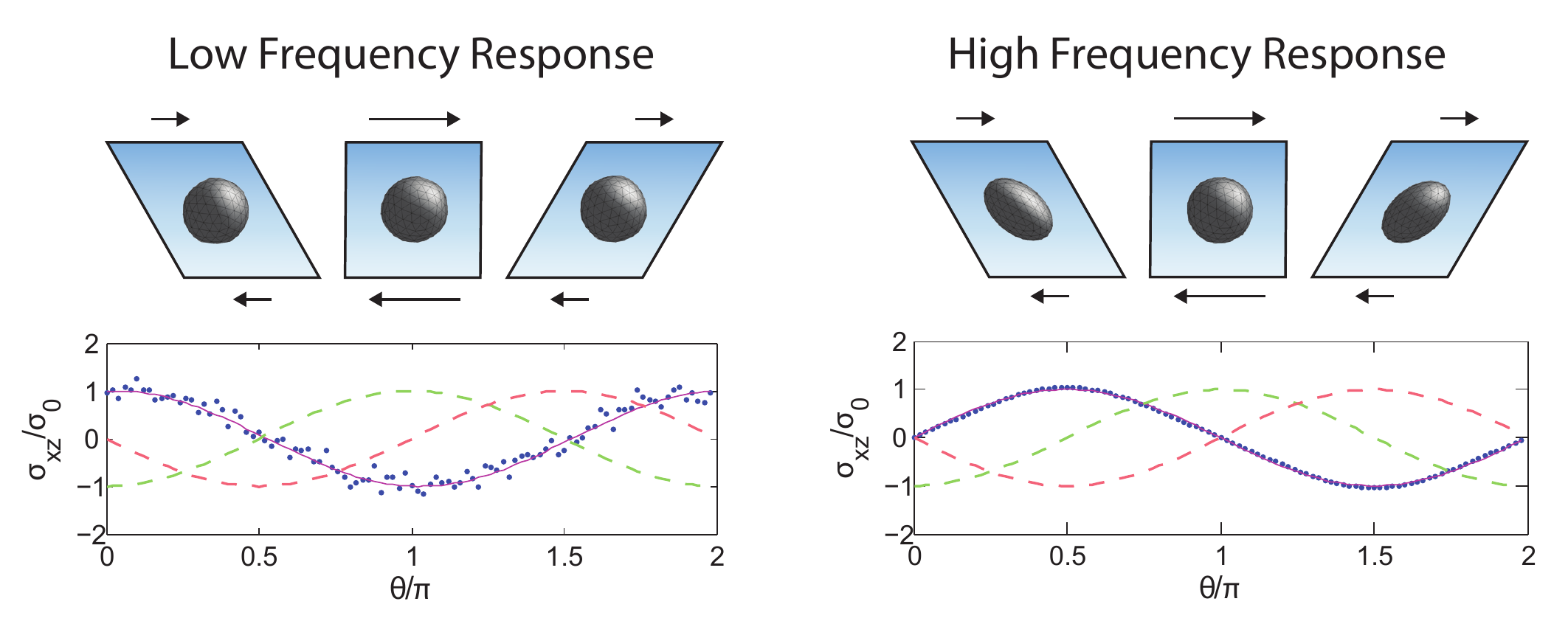}
\caption[Vesicle Response]
{Polymerized Vesicle subject to Oscillatory Shear.
(left) At low frequencies the distortion of the vesicle shape 
is rather small and masked by the 
thermal fluctuations.  When averaging over many cycles, the 
vesicle stress follows closely in phase with the applied 
shear stress plotted in green.  (right) At high frequencies the distortion of the 
vesicle shape is significant and the vesicle stress follows 
closely the strain plotted in red.  The particular low frequency responses shown are for
$\omega =  3.9294 \times 10^{-3} \mbox{ns}^{-1}$, 
$\dot{\gamma} = 1.9647 \times 10^{-3} \mbox{ns}^{-1}$,
$\sigma_0 = 3.7114\times 10^{8} \mbox{amu}\cdot\mbox{nm}^{-1}\cdot\mbox{ns}^{-2}$.
The particular high frequency responses shown are for
$\omega = 1.2426 \times 10^2 \mbox{ns}^{-1}$,
$\dot{\gamma} = 6.2129\times 10^{1} \mbox{ns}^{-1}$,
$\sigma_0 = 4.6314\times 10^{10} \mbox{amu}\cdot\mbox{nm}^{-1}\cdot\mbox{ns}^{-2}$.
The vesicle configurations for each frequency is shown for $\theta = 1.6\pi$, $0.0\pi$ and $0.4\pi$ 
respectively left to right.
}
\label{figure_vesicles_sheared_low_freq}
\label{figure_vesicles_sheared_high_freq}
\end{figure}

We investigate the response of the vesicle to oscillatory shear 
stresses applied with time-varying rate 
$\dot{\gamma} = \dot{\gamma}^0\cos(\omega t)$.
We consider a dilute regime in which it is sufficient
to study a single polymerized vesicle subject to the  
oscillatory shear.  The effective stress tensor associated
with the vesicle suspension at a given time 
$\sigma(t)$ is estimated using the approach 
discussed in Section~\ref{sec_stress_estimator}.

As a measure of the material response, we consider the dynamic complex 
modulus $G(\omega) = G'(\omega) + iG''(\omega)$, whose components are  
defined from measurements of the stress as the least-squares fit 
of the periodic stress component $\sigma_{xz}(t)$ by the function 
$g(t) = G'(\omega) \gamma^0 \cos(\omega t) + G''(\omega) \gamma^0 \sin(\omega t)$.
This offers one characterization of the response of the material 
to oscillating applied shear stresses and strains as the frequency
$\omega$ is varied.  The $G'$ is referred to as the Elastic Storage
Modulus and $G''$ is referred to as the Viscous Loss Modulus.
These dynamic moduli are motivated by considering the linear
response of the stress components $\sigma_{xz}(t)$ to applied 
stresses and strains.  For many materials linearity holds to 
a good approximation over a wide range of frequencies provided 
the amplitudes of the applied stresses and strains are 
not too large~\cite{bird1987}.  

To estimate the dynamic complex modulus in practice, the 
least-squares fit is performed for $\sigma_{xz}(t)$ over the entire
stochastic trajectory of a simulation (after some transient period).
Throughout our discussion we refer to $\theta = \omega{t}$ as the 
phase of the periodic response.  In our simulations, the maximum strain 
over each period was chosen to always be half the periodic unit cell in 
the x-direction, corresponding to a strain amplitude of $\gamma^0 = \frac{1}{2}$.  
This was achieved by adjusting the shear rate amplitude for
each frequency using the expression $\dot{\gamma}^0 = \gamma^0\omega$.

We performed simulations subjecting the vesicle to shear over a 
wide range of frequencies.   At low frequency the 
distortion of the vesicle shape was found to be small 
and masked by thermal fluctuations when averaged over 
hundreds of periods.  At low frequency the vesicle 
stresses appear to have sufficient time to equilibrate to the 
applied shear stresses.  This is manifested in $\sigma_{xz}(t)$, 
which is seen to track very closely the applied stress, 
see Figure~\ref{figure_vesicles_sheared_low_freq}.  
At high frequencies, the vesicle shape was found to become 
visibly distorted and the vesicle stresses did not appear 
to have sufficient time to equilibrate to the applied 
shear stresses.  These distortions can be seen in the 
configurations for phase $\theta = 1.6, 0.4$.  The $\sigma_{xz}$
is seen to be out of phase with the applied stresses but in 
phase with respect to the applied strain, 
see Figure~\ref{figure_vesicles_sheared_high_freq}.  

\begin{figure}[t]
\centering
\includegraphics[width=5in]{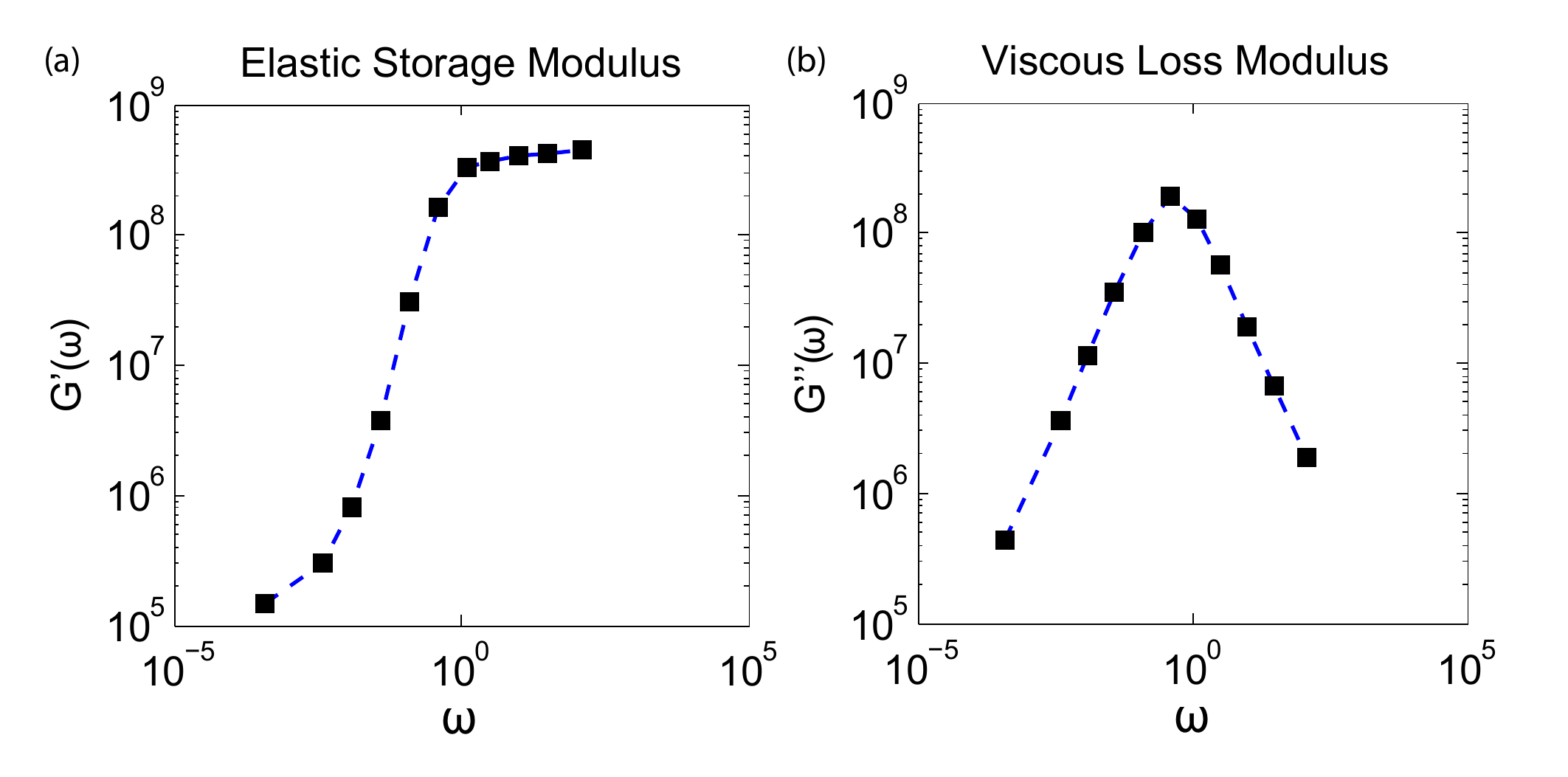}
\caption[Vesicle Response]
{Vesicle Frequency Response : Dynamic Moduli.  The Elastic Storage Modulus $G'(\omega)$
is shown in panel (a) and the Viscous Loss Modulus $G''(\omega)$ is shown in panel (b).  
}
\label{figure_vesicle_complex_modulus}
\end{figure}

These responses can be quantitated by considering the dynamic
moduli.  From the simulations, it is seen for the 
low frequency responses that the vesicle stress follows
closely the applied stress.  It is also found that the 
Viscous Loss Modulus is significantly larger than 
the Elastic Storage Modulus.  For the high frequency
response, it was found that the Elastic Storage Modulus 
increases and is eventually much larger than the 
Viscous Loss Modulus.  It was also found that the 
Viscous Loss Modulus exhibited a non-monotonic behavior at 
intermediate frequencies, see Figure~\ref{figure_vesicle_complex_modulus}.
A description of the parameters and specific values used in 
the simulations can be found in Table~\ref{table_vesicle_param_descr}
and Table~\ref{table_vesicle_param_value}.

\begin{table}[h]
\centering
\begin{tabular}{|l|l|}
\hline
Parameter & Description \\
\hline
$N$                           & Number of mesh points in each direction.\\
$\Delta{x}$                   & Mesh spacing.                  \\
$L$                           & Domain size in each direction. \\
$T$                           & Temperature.                   \\
$k_B$                         & Boltzmann's constant. \\                              
$\mu$                         & Dynamic viscosity of the solvent fluid. \\                             
$\rho$                        & Mass density of the solvent fluid. \\                             
$K_1$                         & Vesicle bond stiffness. \\
$K_2$                         & Vesicle bending stiffness. \\
$D$                           & Vesicle diameter. \\
$\omega$                      & Frequency of oscillating shearing motion. \\
$\theta$                      & Phase of the oscillatory motion, $\theta = \omega t$. \\
$\dot{\gamma}$                & Shear rate. \\
$\dot{\gamma}^0$              & Shear rate amplitude. \\
$\gamma$                      & Strain rate. \\
$\gamma^0$                    & Strain rate amplitude. \\
\hline
\end{tabular}
\caption[Vesicle Parameter Description]
{Description of the parameters used in simulations of the polymerized vesicle.
\label{table_vesicle_param_descr}}
\end{table}

\begin{table}[t]
\centering
\begin{tabular}{|l|l|}
\hline
Parameter & Value \\
\hline
$N$                           & $27$ \\
$\Delta{x}$                   & $7.5 \mbox{ nm}$ \\
$L$                           & $2.025\times10^2 \mbox{ nm}$ \\
$T$                           & $300 \mbox{ K}$ \\
$k_B$                         & $8.3145 \times10^3    \mbox{ nm}^2\cdot\mbox{amu}\cdot\mbox{ns}^{-2}\cdot\mbox{K}^{-1}$ \\
$\mu$                         & $6.0221 \times 10^{5} \mbox{ amu}\cdot\mbox{cm}^{-1}\cdot\mbox{ns}^{-1}$ \\
$\rho$                        & $6.0221 \times 10^{2} \mbox{ amu}\cdot\mbox{nm}^{-3}$ \\
$\mbox{K}_1$                  & $2.2449 \times 10^{7} \mbox{ amu}\cdot \mbox{ns}^{-2}$ \\
$\mbox{K}_2$                  & $8.9796\times 10^{7}$ \\
$D$                           & $50 \mbox{ nm}$ \\
\hline
\end{tabular}
\caption[Vesicle Parameter Description]
{Fixed values of the parameters used in simulations of the polymerized vesicle.
\label{table_vesicle_param_value}}
\end{table}

\clearpage
\newpage

\subsection{Application III: Aging of the Shear Viscosity of a Gel-like Material}

\label{sec_appl_aging_shear_viscosity}

As a further demonstration of how the computational methods 
can be used, we investigate the aging of a gel-like material 
subject to shear.  The methods are used to study how the shear 
viscosity changes over time as the gel is subjected to shear 
at a constant rate.

\begin{figure}[t]
\centering
\includegraphics[width=5in]{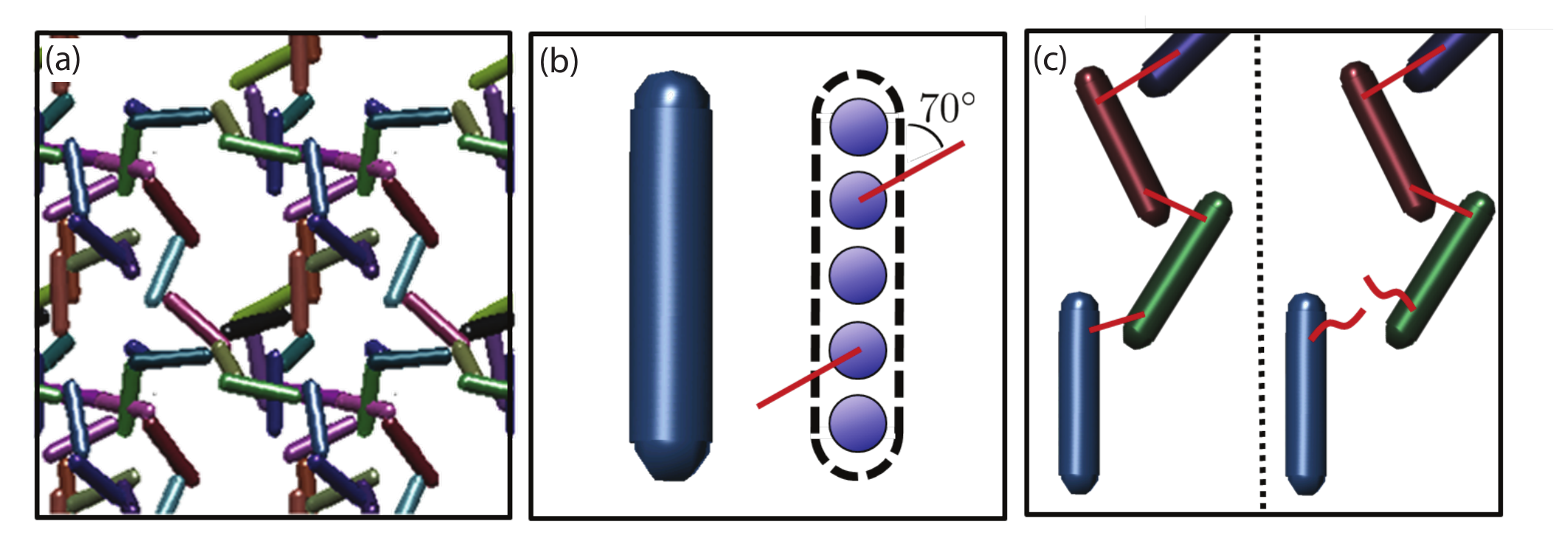}
\caption[Gel Microstructures.]
{Microstructure of the Gel-like Material.  (a) The microstructure
of the gel is comprised of polymeric chains which bind together.  
(b) The polymeric chains are each comprised of 
five control points and each have specialized binding sites 
at the second and fourth control point.
(c) The inter-polymer bonds have a preferred extension and angle.
When an inter-polymer bond is strained beyond $50\%$ of its
preferred rest length, the bond breaks irreversibly.
}
\label{figure_gelSchematic}
\end{figure}

The gel-like material is modeled as a collection of 
polymer chains which are able to bond together 
at two specialized sites along the chain, see Figure~\ref{figure_gelSchematic}. 
The energy associated with the mechanics of the individual
polymer chains and the bonds which they form are given by
\begin{eqnarray}
\phi_{1}(r)       & = & \frac{1}{2}K_{1} (r - r_{0,1})^2 \\
\phi_{2}(\boldsymbol{\tau}_1,\boldsymbol{\tau}_2) 
                  & = &  
\frac{1}{2} K_{2}
\left|\boldsymbol{\tau}_{1} - \boldsymbol{\tau}_{2}\right|^2 \\
\phi_{3}(r)       & = & \sigma^2 K_{3}\exp\left[-\frac{(r - r_{0,3})^2}{2\sigma^2}\right] \\
\phi_{4}(\theta)  & = & -K_{4} \cos(\theta - \theta_{0,4}).
\end{eqnarray}
The $r$ is the separation distance between two control points,
$\theta$ is the bond angle between three control points, 
and $\boldsymbol{\tau}$ is a tangent vector along the polymer 
chain, see Figure ~\ref{figure_gelSchematic}.

The $\phi_1$ energy accounts for stretching of a bond within a polymer chain from its preferred
extension $r_{0,1}$. The $\phi_2$ energy accounts for bending of the polymer chain locally.
To account for interactions at the specialized binding sites of the polymers the potentials
$\phi_3$ and $\phi_4$ are introduced.  The potential $\phi_3$ gives the energy of the 
bond between the two polymer chains and penalizes deviation from the preferred bond extension
$r_{0,3}$.  The exponential of $\phi_3$ is introduced so that the resistance in the 
bond behaves initially like a harmonic bond but decays rapidly to zero when the bond is 
stretched beyond the length $\sigma$.  The potential $\phi_4$ 
gives the energy for the preferred bond angle when two of the polymer 
chains are bound together.

The total energy of the system is given by 
\begin{eqnarray}
\Phi[\mathbf{X}] & = & E_1[\mathbf{X}] + E_2[\mathbf{X}] + E_3[\mathbf{X}] + E_4[\mathbf{X}] \\
E_1[\mathbf{X}]  & = & \sum_{(i,j)   \in \mathcal{Q}_1}  \phi_1(r_{ij}), \mbox{\hspace{0.25cm}}
E_2[\mathbf{X}]   =  \sum_{(i,j,k) \in \mathcal{Q}_2} \phi_2(\boldsymbol{\tau}_{ij},\boldsymbol{\tau}_{jk}) \\
E_3[\mathbf{X}]  & = & \sum_{(i,j)   \in \mathcal{Q}_3}  \phi_3(r_{ij}), \mbox{\hspace{0.25cm}}
E_4[\mathbf{X}]   =  \sum_{(i,j,k) \in \mathcal{Q}_4} \phi_4(\theta_{ijk}).
\end{eqnarray}
The sets $\mathcal{Q}_k$ define the interactions according to the 
structure of the individual polymer chains and the topology of 
the gel network.  When bonds are stretched beyond the critical length $3\sigma$ they 
are broken irreversibly, which results in the sets $\mathcal{Q}_3$ and $\mathcal{Q}_4$
being time dependent.

\begin{figure}[t]
\centering
\includegraphics[width=4in]{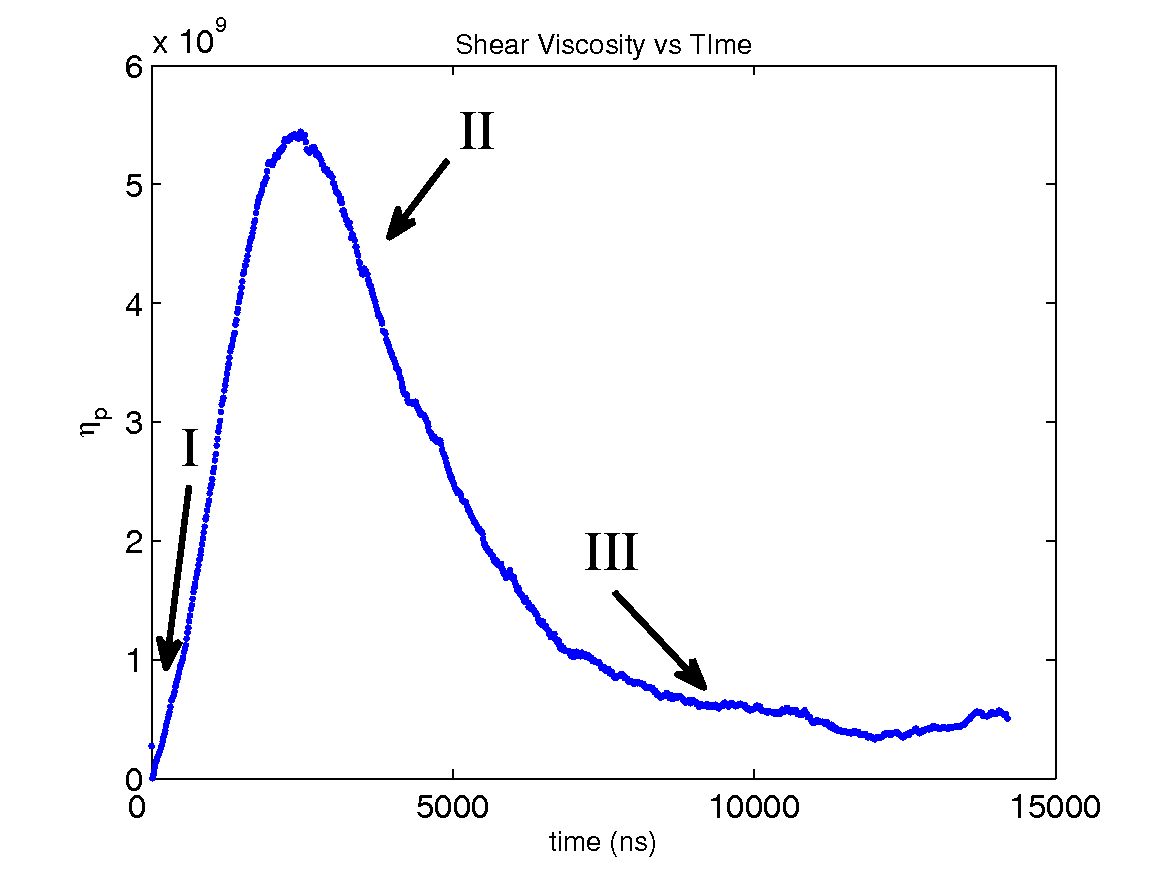}
\caption[Shear Viscosity of a Gel-like Material]
{Aging of the Shear Viscosity Over Time.  The aging of
the shear viscosity exhibits roughly three stages,
labeled by I, II, III.  
In the first stage, the gel-network maintains its 
integrity.  Contributions to the shear viscosity 
arise from stretching of the inter-polymer and 
intra-polymer bonds.  In the second stage, the 
inter-polymer bonds of the gel-network break.
The polymers are then free to align with the direction of 
shear which results in relaxation of the intra-polymer 
bonds to their preferred rest-length.
In the third stage, the contributions to the 
shear viscosity arise from thermal fluctuations 
that drive transient misalignments of the polymers 
with the direction of shear.  These stages are each discussed in 
more detail in Section~\ref{sec_appl_aging_shear_viscosity}.
}
\label{figure_thixotropyGelLikeMaterial}
\end{figure}

To study the rheological response of the gel-like material 
the system is subjected to shear at a constant rate.  
To obtain an effective macroscopic stress $\sigma_p$ 
for the system, we use the approach from 
Section~\ref{sec_stress_estimator}.
To characterize the rheology of the gel,
we consider the shear viscosity defined by 
\begin{eqnarray}
\eta_p & = & {\sigma_p^{(s,v)}}/{\dot{\gamma}}.
\end{eqnarray}
The $\dot{\gamma}$ is the rate of the applied shear.
In the notation, the superscript $(s,v)$ indicates the 
tensor component with the index $s$ corresponding to the 
direction of shear and the index $v$ corresponding to 
the direction of shear induced velocity.  The contributions of 
the solvent fluid to the shear viscosity are assumed to be 
Newtonian and can be considered separately~\cite{Bird1987Vol2}.

To investigate how the shear viscosity of the gel behaves
over time, multiple simulations were performed starting
from an undisturbed configuration of the gel network.
A shear was then applied to the unit cell boundary to
induce a shear deformation of the network.  The
shear deformation over time resulted in the breakage 
of bonds of the gel network.  To investigate 
how the macroscopic material properties depend on
the reorganization of the polymer chains, 
we considered the shear viscosity over time.

\begin{figure}[t]
\centering
\includegraphics[width=5in]{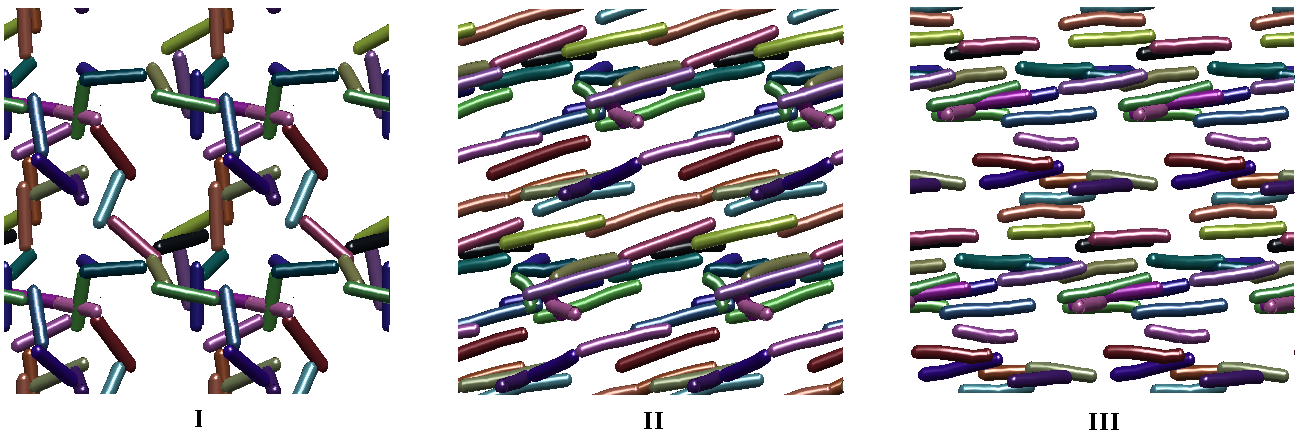}
\caption[Microstructure of the Gel-like Material]
{Microstructure Configurations for the Gel-like Material.
Configurations of the microstructure are shown as the 
gel ages for each of the three stages discussed in 
Section~\ref{sec_appl_aging_shear_viscosity}.
The times shown in these figures 
from left to right are $t = 0\mbox{ ns}$, $t = 2844\mbox{ ns}$, $t = 7111\mbox{ ns}$.
}
\label{figure_thixotropyGelLikeMaterial_configComposite}
\end{figure}

From the simulations, an interesting behavior was 
found.  The material initially exhibited an 
increased shear viscosity before eventually 
settling down to a steady-state value.  It was
found that the responses of the gel-like material 
to shear can be roughly divided into three stages.
In the first, there is an initial increase of the 
shear viscosity which can be attributed to the 
stretching of the inter-chain bonds between 
the polymer chains and the intra-chain bonds 
within each polymer chain.  This occurs as the 
gel as a whole is strained for a relatively 
short period.  The bonds between the polymer 
chains are observed to break with the remaining 
contributions to the stress arising from the 
shear stresses of the fluid which stretch the 
polymer chains.  The shear viscosity
in this stage and the microstructure of the gel
are shown in the regions labeled by I in 
Figure~\ref{figure_thixotropyGelLikeMaterial} 
and Figure~\ref{figure_thixotropyGelLikeMaterial_configComposite}.

In the second stage, the individual polymer chains 
are seen to rotate and to start aligning with the 
direction of the shear.  As a result of the intra-chain 
restoring forces the strain of the individual polymer chains
is also seen to relax.  The increased alignment and reduction
of strain of the polymer chains yields an overall 
decrease in the forces transmitted in the direction 
of the shear gradient.  Consequently, the shear viscosity 
begins to decrease, see the regions labeled by 
II in Figure~\ref{figure_thixotropyGelLikeMaterial} 
and Figure~\ref{figure_thixotropyGelLikeMaterial_configComposite}.

In the third, and last stage, the chains eventually 
settle into a statistical steady-state in which the 
thermal fluctuations drive the chains to 
misalign transiently with the direction of shear.  
These misaligned excursions by the polymer chains
result on average in non-negligible forces transmitted 
in the direction of the shear gradient.  As a consequence, 
the shear viscosity has a non-zero steady-state value.
This is shown in the regions labeled by III in 
Figure~\ref{figure_thixotropyGelLikeMaterial} 
and Figure~\ref{figure_thixotropyGelLikeMaterial_configComposite}.
Similar aging phenomenon is seen for many different types of 
soft materials and complex fluids and is often referred to as 
``thixotropy'', see~\cite{Barnes1997,Bird1987Vol2,Doi1986}.
For the specific physical parameters used in these simulations 
see Table \ref{table_gel_param_descr} and \ref{table_gel_param_value}.

\begin{table}[t]
\centering
\begin{tabular}{|l|l|}
\hline
Parameter & Description \\
\hline
$N$                             & Number of mesh points in each direction.\\
$\Delta{x}$                     & Mesh spacing.                  \\
$\Delta{t}$                     & Time step.                     \\
$L$                             & Domain size in each direction. \\
$T$                             & Temperature.                   \\
$k_B$                           & Boltzmann's constant. \\                              
$\mu$                           & Dynamic viscosity of the solvent fluid. \\                             
$\rho$                          & Mass density of the solvent fluid. \\                             
$\dot{\gamma}$                  & Shear rate. \\
$N_p$                           & Number of polymer chains. \\
$N_s$                           & Number of control points per polymer chain. \\
$r_p$                           & Polymer effective cylindrical radius. \\
$K_{1}$                         & Stiffness of the bonds of the polymer chain. \\
$r_{0,1}$                       & Rest length of the bonds of the polymer chain. \\
$K_{2}$                         & Bending stiffness of the polymer chain. \\
$K_{3}$                         & Stiffness of the bonds at a polymer binding site. \\
$r_{0,3}$                       & Rest length of the bond at a polymer binding site. \\
$K_{4}$                         & Bending stiffness of the bond at a polymer binding site. \\
$\theta_{0,4}$                  & Preferred angle of a bond at a polymer binding site. \\
\hline
\end{tabular}
\caption[Gel-Like Material Parameter Description]
{Description of the parameters used in simulations of the gel-like material.
\label{table_gel_param_descr}}
\end{table}

\begin{table}[t]
\centering
\begin{tabular}{|l|l|}
\hline
Parameter & Value \\
\hline
$N$                           & $72$ \\
$\Delta{x}$                   & $11.25  \mbox{ nm}$ \\
$\Delta{t}$                   & $1.4222 \mbox{ ns}$ \\
$L$                           & $810 \mbox{ nm}$ \\
$T$                           & $300 \mbox{ K}$ \\
$k_B$                         & $8.3145 \times10^3    \mbox{ nm}^2\cdot\mbox{amu}\cdot\mbox{ns}^{-2}\cdot\mbox{K}^{-1}$ \\
$\mu$                         & $6.0221 \times 10^{5} \mbox{ amu}\cdot\mbox{cm}^{-1}\cdot\mbox{ns}^{-1}$ \\
$\rho$                        & $6.0221 \times 10^{2} \mbox{ amu}\cdot\mbox{nm}^{-3}$ \\
$\dot{\gamma}$                & $1.2 \times 10^{-3} \mbox{ ns}^{-1}$ \\
$N_p$                         & $110$ \\
$N_s$                         & $5$ \\
$r_p$                         & $15 \mbox{ nm}$ \\
$K_{1}$                       & $2.9932 \times 10^{5} \mbox{ amu}\cdot\mbox{ns}^{-2}$  \\
$r_{0,1}$                     & $30 \mbox{ nm}$ \\
$K_{2}$                       & $2.9932 \times 10^{8}$ \\
$K_{3}$                       & $2.9932 \times 10^{5} \mbox{ amu}\cdot\mbox{ns}^{-2}$  \\
$r_{0,3}$                     & $30 \mbox{ nm}$ \\
$K_{4}$                       & $2.9932 \times 10^{8}$ \\
$\theta_{0,4}$                & $70\,^{\circ}$ \\
\hline
\end{tabular}
\caption[Gel-like Material Parameter Description]
{Fixed values of the parameters used in simulations of the gel-like material.
\label{table_gel_param_value}}
\end{table}

The modeling approach presented here, along with the 
computational methods, allow for other types of
phenomena to be studied.  This includes looking at 
the case in which the bonds between the polymer 
chains are able to reform.  An interesting 
investigation in this case would be to study 
the kinetics and organization of the
gel network when subject to both bond 
breaking and bond formation when the shear 
is greatly decreased or arrested for 
periods of time.  

The computational methods
also provide a straight-forward means 
to incorporate active forces operating 
on the filaments of the gel network.  
This could be useful in the study of 
biological materials, such as actin 
where motor proteins act like active 
cross-linkers to slide filaments
past one another~\cite{Danuser2006,Li2005,Mizuno2007}.  
These effects could readily be taken 
into account in such simulations.   

\clearpage
\newpage

\section{Conclusions}
We have presented an approach for incorporating shear into fluctuating hydrodynamics methods for 
studies of the rheological responses of complex fluids and soft materials.  We have shown how
generalized periodic boundary conditions and stochastic numerical methods can be formulated 
to handle shear.  We have shown that formulating the momentum conservation equations 
in a moving frame of reference that tracks the distortion of the unit cell provides a number
of advantages.  We have furthermore presented some analysis of the time-dependent discretization 
methods to show how stochastic driving fields can be introduced to satisfy a 
fluctuation-dissipation balance despite truncation errors.  As can be seen from the 
example applications presented, the introduced methods provide a number of ways to simulate phenomena 
relevant to rheological responses.  We expect the presented approaches to be useful 
in adopting fluctuating hydrodynamics descriptions to investigate models and 
phenomena relevant in studies of complex fluids and soft materials.  

\section{Software}
A related simulation package for the discussed methods can be found at 
\url{https://github.com/atzberg/mango-selm}.  This includes an interface to
readily setup models and perform simulation studies. Additional tutorials, 
videos, and other information can be found at \url{http://atzberger.org/}.

\section{Acknowledgements}
The author P.J.A. acknowledges support from research grant 
NSF CAREER DMS - 0956210 and DOE CM4.
We would especially like to thank Phil Pincus, Aleksandar Donev, 
Alejandro Garcia, John Bell, and Tony Ladd for stimulating 
conversations about this work.  This paper is dedicated in 
memorial to Tom Bringley, whose passion for life, mathematics, 
and science was an inspiration to all who knew him.

\bibliographystyle{siam}
\bibliography{paperBibliography}

\appendix

\section{A Fluctuation-Dissipation Principle for Time-Dependent Operators}
\label{appendix_fluct_dissip}
Consider the stochastic process given by
\begin{eqnarray}
d\mathbf{z}_t & = & L(t)\mathbf{z}_t dt + Q(t)d\mathbf{B}_t \\
G(t)          & = & QQ^T.
\end{eqnarray}
We now establish the following fluctuation-dissipation relation
\begin{eqnarray}
\label{equ_G_t_fluct_dissp}
G(t) = -L(t)\bar{C} - \bar{C}^TL(t)^T.
\end{eqnarray}
This relates the covariance $G(t)$ of the stochastic 
driving field to a time-dependent dissipative operator 
$L(t)$ and a time-independent equilibrium covariance 
$\bar{C}$.  We show that this relation allows for 
$G(t)$ to be chosen to ensure that the stochastic 
dynamics exhibits at statistical steady-state 
equilibrium fluctuations with the specified
covariance $\bar{C}$.

Let the covariance at time $t$ be denoted by
\begin{eqnarray}
C(t) = \langle\mathbf{u}(t)\mathbf{u}(t)^T\rangle.
\end{eqnarray}
By Ito's Lemma the second moment satisfies
\begin{eqnarray}
\label{equ_M_t}
dC(t) = \left(L(t)C(t) + C(t)^TL(t)^T + G(t)\right) dt.
\end{eqnarray}
It will be convenient to express this equation 
by considering all of the individual entries of the 
matrix $C(t)$ collected into a single column vector 
denoted by $\mathbf{c}_t$.  Similarly, for covariance 
matrix $G(t)$ we denote the column vector of entries
by $\mathbf{g}_t$ and for $\bar{C}$ by 
$\bar{\mathbf{c}}$.  Since the products $L(t)C(t)$
and $C(t)^TL(t)^T$ are both linear operations in 
the entries of the matrix $C(t)$ we can express 
this in terms of multiplication by of a matrix 
$A(t)$ acting on $\mathbf{c}_t$.  

This notation allows for equation~\ref{equ_M_t} to be 
expressed equivalently as
\begin{eqnarray}
\label{equ_m_t}
d\mathbf{c}_t = \left(A(t) \mathbf{c}_t + \mathbf{g}_t\right) dt.
\end{eqnarray}
The equation~\ref{equ_M_t} can be solved formally  
by the method of integrating factors to obtain
\begin{eqnarray}
\label{equ_m_t_int}
\mathbf{c}_t = e^{\Xi(0,t)}\mathbf{c}_0
+ \int_0^t e^{\Xi(s,t)} \mathbf{g}_s ds
\end{eqnarray}
where $\Xi(s,t) = \int_s^t A(r) dr$.

The fluctuation-dissipation relation given by 
equation~\ref{equ_G_t_fluct_dissp} is equivalent to choosing 
\begin{eqnarray}
\mathbf{g}_s = -A(s)\bar{\mathbf{c}}.
\end{eqnarray}
For this choice, a useful identity is 
\begin{eqnarray}
e^{\Xi(s,t)} \mathbf{g}_s = \frac{\partial}{\partial s} e^{\Xi(s,t)}\bar{\mathbf{c}}.
\end{eqnarray}
Substitution into equation~\ref{equ_m_t_int} gives
\begin{eqnarray}
\label{equ_m_t_int2}
\mathbf{c}_t = e^{\Xi(0,t)}\mathbf{c}_0
+ \left(e^{\Xi(t,t)} - e^{\Xi(0,t)}\right) \bar{\mathbf{c}}.
\end{eqnarray}

Now, if $L(t)$ is negative definite uniformly in time, 
$\mathbf{v}^T L(t) \mathbf{v} < \alpha_0 < 0$,
then $A(t)$ is also uniformly negative definite.
This implies that 
\begin{eqnarray}
\label{equ_lim_e_Xi}
\lim_{t\rightarrow \infty} e^{\Xi(0,t)} = 0.
\end{eqnarray}
Taking the limit of both sides of equation~\ref{equ_m_t_int2}
and using equation~\ref{equ_lim_e_Xi} yields
\begin{eqnarray}
\label{equ_m_t_limit}
\lim_{t\rightarrow \infty} \mathbf{c}_t 
= \mathbf{\bar{c}}.
\end{eqnarray}

This shows that the stochastic driving field with 
covariance given by equation~\ref{equ_G_t_fluct_dissp} 
yields equilibrium fluctuations with covariance $\bar{C}$.
This extends the fluctuation-dissipation relation
to the case of time-dependent operators.

\section{Coupling Kernel : Peskin $\delta_a$-Function }
\label{appendix_delta_func}
We utilize a similar kernel function to
the one which was used in the 
Immersed Boundary Method 
to represent structures~\cite{Atzberger2007a, Peskin2002}.  
This is given by
\begin{eqnarray}
\label{equ_phi_def}
\phi(r) & = & \left\{
\begin{array}{ll}
0
                  & \mbox{, if $r \leq -2$} \\
                  & \\
\frac{1}{8} \left(5 + 2r - \sqrt{-7 - 12r - 4r^2} \right) 
                  & \mbox{, if $-2 \leq r \leq -1$} \\
                  & \\
\frac{1}{8} \left(3 + 2r + \sqrt{1 - 4r - 4r^2} \right) 
                  & \mbox{, if $-1 \leq r \leq 0$} \\
                  & \\
\frac{1}{8} \left(3 - 2r + \sqrt{1 + 4r - 4r^2} \right) 
                  & \mbox{, if $0 \leq r \leq 1$} \\
                  & \\
\frac{1}{8} \left(5 - 2r - \sqrt{-7 + 12r - 4r^2} \right) 
                  & \mbox{, if $1 \leq r \leq 2$} \\
                  & \\
0
                  & \mbox{, if $2 \leq r$.} 
                  \\
\end{array}
\right.
\end{eqnarray}
For three dimensional systems the function 
$\delta_a$ is given by
\begin{eqnarray}
\label{equ_phi_def_delta_a}
\delta_a(\mathbf{r}) = \frac{1}{a^3}
\phi\left(\frac{\mathbf{r}^{(1)}}{a}\right)
\phi\left(\frac{\mathbf{r}^{(2)}}{a}\right)
\phi\left(\frac{\mathbf{r}^{(3)}}{a}\right), 
\end{eqnarray}
where the superscript indicates the index of the vector component.
To maintain good numerical properties, the particles are 
restricted to sizes $a = n\Delta{x}$, where $n$ is a positive 
integer.  For a derivation and a detailed discussion of 
the properties of this kernel function, 
see~\cite{Peskin2002, Atzberger2007a, Bringley2008}.

\section{Table}
\vspace{0.5cm}
\label{appendix_rough_calc}

\begin{tabular}{|l|l|}
\hline
Parameter & Description \\
\hline
$N_A$                         & Avogadro's number. \\
\mbox{amu}                    & Atomic mass unit. \\
\mbox{nm}                     & Nanometer. \\
\mbox{ns}                     & Nanosecond. \\
$k_B$                         & Boltzmann's Constant. \\
$T$                           & Temperature. \\
$\eta$                        & Dynamic viscosity of water. \\
$\gamma_s = 6\pi \eta R$        & Stokes' drag of a spherical particle. \\
\hline
\end{tabular}
\vspace{1cm}

\begin{tabular}{|l|l|}
\hline
Parameter & Value \\
\hline
$N_A$                         & $6.02214199\times 10^{23}$. \\
\mbox{amu}                    & $1/10^3 N_A$ \mbox{kg}. \\
\mbox{nm}                     & $10^{-9}\hspace{0.1cm} \mbox{m}$. \\
\mbox{ns}                     & $10^{-9}\hspace{0.1cm} \mbox{s}$. \\
$k_B$                         & $8.31447\times 10^{3} \hspace{0.1cm} \mbox{amu}\hspace{0.1cm}\mbox{nm}^2/\mbox{ns}^2\hspace{0.1cm}K$. \\
$T$                           & $300\mbox{K}$. \\
$\eta$                        & $6.02214199\hspace{0.1cm} \mbox{amu}/\mbox{cm}\hspace{0.1cm}\mbox{ns}$. \\
\hline
\end{tabular}
\vspace{1cm}

\end{document}